\documentclass[conference]{IEEEtran}
\usepackage{cite}
\usepackage{amsmath,amssymb,amsfonts}
\usepackage{algorithmic}
\usepackage{graphicx}
\usepackage{textcomp}
\usepackage[hyphens]{url}
\usepackage{booktabs}
\usepackage{multirow}
\usepackage{multicol}
\usepackage{graphicx} % omit 'demo' option in real document
\usepackage{subfig}
\usepackage[table,xcdraw]{xcolor}
\usepackage[symbol]{footmisc}
\renewcommand{\thefootnote}{\fnsymbol{footnote}}
\newcommand{\ieOne}{multi resolution hashgrid~}
\newcommand{\ieTwo}{multi resolution densegrid~}
\newcommand{\ieThree}{low resolution densegrid~}
\newcommand{\iEOne}{Multi resolution hashgrid~}
\newcommand{\iETwo}{Multi resolution densegrid~}
\newcommand{\iEThree}{Low resolution densegrid~}
\newcommand{\ieone}{MRHG~}
\newcommand{\ietwo}{MRDG~}
\newcommand{\iethree}{LRDG~}

\newcommand{\iePercentHashgrid}{40.24\%~}
\newcommand{\iePercentDensegrid}{24.63\%~}
\newcommand{\iePercentDensegridLowRes}{24.15\%~}
\newcommand{\mlpPercentHashgrid}{32.13\%~}
\newcommand{\mlpPercentDensegrid}{35.37\%~}
\newcommand{\mlpPercentDensegridLowRes}{35.81\%~}

\newcommand{\ieOneIEPercent}{40.24}
\newcommand{\ieOneMLPPercent}{32.12}
\newcommand{\ieOneIEMLP}{72.37}
\newcommand{\ieTwoIEPercent}{24.63}
\newcommand{\ieTwoMLPPercent}{35.37}
\newcommand{\ieTwoIEMLP}{60.0}
\newcommand{\ieThreeIEPercent}{24.15}
\newcommand{\ieThreeMLPPercent}{35.37}
\newcommand{\ieThreeIEMLP}{59.96}

\newcommand{\nG}{neural graphics~}
\newcommand{\IE}{input encoding~}
\newcommand{\MLP}{multi-layer perceptron~}
\newcommand{\MLPs}{multi-layer perceptrons~}
\newcommand{\NFP}{neural fields processor~}
\newcommand{\nfp}{NFP~}
\newcommand{\NPC}{neural graphics processing cluster~}
\newcommand{\npc}{NGPC}

\newcommand{\npcOAvgHG}{12.94}
\newcommand{\npcTAvgHG}{20.85}
\newcommand{\npcFAvgHG}{33.73}
\newcommand{\npcEAvgHG}{39.04}
\newcommand{\maxperf}{58.36}
\newcommand{\fuserest}{9.94}

\newcommand{\npcOAvgDG}{9.05}
\newcommand{\npcTAvgDG}{14.22}
\newcommand{\npcFAvgDG}{22.57}
\newcommand{\npcEAvgDG}{26.22}

\newcommand{\npcOAvgLR}{9.37}
\newcommand{\npcTAvgLR}{14.66}
\newcommand{\npcFAvgLR}{22.97}
\newcommand{\npcEAvgLR}{26.4}

\newcommand{\appN}{NeRF~}
\newcommand{\appS}{NSDF~}
\newcommand{\appV}{NVR~}
\newcommand{\appI}{GIA~}

\newcommand{\percentOarea}{4.52}
\newcommand{\percentOpow}{2.75}
\newcommand{\percentTarea}{9.04}
\newcommand{\percentTpow}{5.51}
\newcommand{\percentFarea}{18.01}
\newcommand{\percentFpow}{11.03}
\newcommand{\percentEarea}{36.18}
\newcommand{\percentEpow}{22.06}

\newcommand{\timeNHG}{231}
\newcommand{\timeSHG}{27.87}
\newcommand{\timeVHG}{6.32}
\newcommand{\timeIHG}{2.12}

\newcommand{\gapNHG}{4625}
\newcommand{\gapNHGOOM}{4}
\newcommand{\gapSHG}{557}
\newcommand{\gapSHGOOM}{3}
\newcommand{\gapIHG}{42.54}
\newcommand{\gapIHGOOM}{2}
\newcommand{\gapVHG}{126.56}
\newcommand{\gapVHGOOM}{2}

\newcommand{\gapNHGOOMVR}{3}
\newcommand{\gapSHGOOMVR}{2}
\newcommand{\gapIHGOOMVR}{1}
\newcommand{\gapVHGOOMVR}{1}

\newcommand{\gapfpsNHGF}{55.50} %updated to 60fps
\newcommand{\gapfpsSHGF}{6.68}
\newcommand{\gapfpsVHGF}{1.51}

\newcommand{\timeSDG}{17.57}
\newcommand{\timeVDG}{4.73}
\newcommand{\timeIDG}{1.48}

\newcommand{\timeSLR}{18.88}
\newcommand{\timeVLR}{5.05}
\newcommand{\timeILR}{1.33}

\newcommand{\hgIeIndv}{246}
\newcommand{\hgMlpIndv}{1232}
\newcommand{\dgIeIndv}{379}
\newcommand{\dgMlpIndv}{1070}
\newcommand{\srIeIndv}{2353}
\newcommand{\srMlpIndv}{1451}

\def\BibTeX{{\rm B\kern-.05em{\sc i\kern-.025em b}\kern-.08em
    T\kern-.1667em\lower.7ex\hbox{E}\kern-.125emX}}

\newcommand{\iscasubmissionnumber}{626}
\title{Hardware Acceleration of Neural Graphics}
\author{\normalsize{ISCA 2023 Submission
    \textbf{\#\iscasubmissionnumber} -- Confidential Draft -- Do NOT Distribute!!}}

\author{Muhammad Husnain Mubarik*\thanks{*Part of the 
work was done during Mubarik's internship at Intel.} \\
mubarik3@illinois.edu \\ UIUC
\and Ramakrishna Kanungo \\
kanungo3@illinois.edu \\ UIUC
\and Tobias Zirr \\
tobias.zirr@intel.com \\ Intel Corporation
\and Rakesh Kumar \\
rakeshk@illinois.edu \\ UIUC
}
\IEEEoverridecommandlockouts
\begin{document}
\maketitle
\thispagestyle{plain}
\pagestyle{plain}

%%%%%% -- PAPER CONTENT STARTS-- %%%%%%%%

\begin{abstract}
% The fundamental  goal of the computer graphics is synthesizing photo-realistic and controllable imagery is
\let\thefootnote\relax\footnotetext{Accepted at the 50th International Symposium on Computer Architecture (ISCA-50)}
% The fundamental  goal of the computer graphics is synthesizing photo-realistic and controllable imagery is
%\let\thefootnote\relax\footnotetext{Accepted at the 50th International Symposium on Computer Architecture (ISCA-50)}

%Rendering and inverse rendering algorithms 
%that drive conventional computer graphics
%have recently shown to be superseded by 
%neural representations (NR) that can handle
%the complete pipeline from sensing to pixels.
%\textcolor{red}{
Rendering and inverse rendering techniques have recently attained powerful new capabilities and building blocks in the form of neural representations (NR), with derived rendering techniques quickly becoming indispensable tools next to classic computer graphics algorithms, covering a wide range of functions throughout the full pipeline from sensing to pixels.
%NRs have recently been used
%to learn the geometric and the material properties of
%the scenes and use the information
%to synthesize photo-realistic imagery, 
%thereby promising a replacement 
%for traditional rendering algorithms
%with scalable 
NRs have recently been used to directly learn the geometric and appearance properties of scenes that were previously hard to capture, and to re-synthesize photo realistic imagery based on this information, 
thereby promising simplifications and replacements for several complex traditional computer graphics problems and algorithms with scalable
quality and predictable performance. 
In this work we ask the question:
{\em Does neural graphics (graphics based on NRs) need hardware support?}
We studied four representative \nG applications (NeRF, NSDF, NVR, and GIA)
showing that, 
if we want to render 4k resolution 
frames at 60 frames per second (FPS) there is a gap of 
$\sim\gapfpsVHGF \times$ to $\gapfpsNHGF \times$
in
the desired performance on current GPUs.
For AR and VR applications, 
there is an even larger gap of
$\sim$ 2-4 orders of magnitude (OOM)
between
the desired performance and the required system power.
We 
identify that the \IE and the \MLP kernels are the
performance bottlenecks, consuming
$\ieOneIEMLP \%$, $\ieTwoIEMLP \%$ and $\ieThreeIEMLP \%$ of application time 
for {\em \ieOne encoding}, {\em \ieTwo encoding} and {\em \ieThree encoding}, respectively. 
We propose a \NPC (\npc{}) -- a scalable and flexible hardware architecture that directly accelerates the \IE and \MLP 
kernels through dedicated engines and supports a wide range of \nG applications.
%We also accelerate the rest of the kernels by fusing them 
%together in Vulkan~\cite{vulkan}, which leads to
%$\sim\fuserest\times$ kernel-level performance improvement compared to 
%Nvidia's "un-fused" implementation~\cite{muller2022instant} of the pre-processing and the post-processing kernels.
%\textcolor{red}{
To achieve good overall application level performance improvements, we also accelerate the rest of the kernels by fusion into a single kernel, leading to a $\sim\fuserest\times$ speedup compared to previous optimized implementations~\cite{muller2022instant} which is sufficient to remove this performance bottleneck.
%}
Our results show that,
\npc{} gives up to $\maxperf\times$ 
end-to-end application-level performance improvement, 
for {\em \ieOne encoding} on average across the four \nG applications, 
the performance benefits are
$\npcOAvgHG\times$, $\npcTAvgHG\times$, $\npcFAvgHG\times$ and 
$\npcEAvgHG\times$ 
%\textcolor{red}{
for the hardware scaling factor of 8, 16, 32 and 64, respectively.
%}
Our results show that with \ieOne encoding, 
\npc{} enables the rendering of 
4k Ultra HD resolution frames at 30 FPS for NeRF and
8k Ultra HD resolution frames at 120 FPS for all our 
other neural graphics applications. 

\end{abstract}

\section{Introduction}
The fundamental goal of classical computer 
graphics is to synthesize 
photo-realistic and controllable 
imagery.
%Over decades, a vast number of algorithms 
%have been 
%developed to achieve this objective. 
%These algorithms can generally be classified as 
%rendering or inverse rendering algorithms.
Rendering algorithms 
%are used
%objective is 
%to 
synthesize an image of a scene 
from the geometric and material 
properties of the scene, often through
%These methods typically 
tracing the path of a photon 
from light source to the 
object and utilizing the information 
about the geometry and scattering 
distributions of the object to simulate the 
interaction 
of light with the object. 
%These
%physics based rendering algorithms 
%require 
%storing and manipulating the physical parameters 
%of the scene 
%such as geometry, texture, reflectivity, 
%opacity, camera parameters and 
%illumination sources etc. 
%In inverse rendering algorithms, 
%the objective is to estimate the 
%physical and material properties of the 
%scene from 
%the existing observations of 
%the scene such as images and videos. 
Inverse rendering algorithms follow the reverse process of rendering. From the final image, they provide guidance about how geometry and materials of a scene need to be adjusted.
%where observations of the scene such as 
%images and videos are used to 
%derive the physical and material properties
%of the scene 
%such as geometry, texture, reflectivity, 
%opacity and 
%illumination etc.  
%For synthesizing the photo-realistic 
%and controllable imagery of a real-world scene, 
%the 
Both rendering and inverse rendering 
algorithms are well-known to be computationally challenging tasks~\cite{ward2007surve,nicolet2021large,shi2008gpu}.
As such, search has gone on for decades to build efficient rendering and inverse rendering algorithms~\cite{meissner2000practical,wittenbrink1998survey,patow2003survey,marschner1998inverse,nicolet2021large}.
\\

%A well known observation is that
%\textcolor{red}{
Since visual data, which is the output of the 
classical rendering and inverse rendering 
algorithms, is usually resilient to approximations~\cite{nepal2014abacus,moons2016energy},
and
%however, 
as neural networks are considered to be 
particularly good function
approximation algorithms,
%} 
a natural question arises: 
{\em can neural networks be used to approximate the 
algorithms used in 
classical computer graphics?}
Recent works~\cite{mildenhall2021nerf, martin2021nerf, muller2022instant,muller2021real,barron2021mip, pumarola2021d,neff2021donerf} 
have shown that such neural representations 
can in fact be superior in learning and representing the 
physical and the material properties of the 
scenes.
Then the information learned in these 
networks is extremely compact and can be used to synthesize 
the photo-realistic imagery.
This process of approximating 
entire or parts of computer graphics using neural networks is known 
as {\em neural graphics}~\cite{tewari2020state}.  Neural graphics promises a fast, deterministic time replacement for traditional rendering algorithms.

\iffalse
Neural graphics is
a rapidly emerging field in recent 
years. 
Just as classical computer graphics, 
the goal of the neural graphics is also 
to synthesize the controllable 
and photo-realistic images/videos. 
The neural graphics  
combines 
the recent advances 
in machine learning and the  
classical computer graphics algorithms
to enable novel ways to synthesize 
controllable and photo-realistic imagery, potentially 
in real-time~\cite{}. \\
\fi

In this paper we ask the question: {\em does \nG need hardware 
support?} 
%We focus on four representative \nG applications: 
%\textcolor{red}{
We focus on four neural graphics applications recently presented as a representative set of neural graphics representations~\cite{muller2022instant}:
%}
1) Neural radiance and density fields (NeRF),
2) Neural signed distance functions (NSDF),
3) Gigapixel image approximation (GIA) and 
4) Neural volume rendering (NVR).
These four applications cover a wide range of
graphics tasks including rendering, novel view synthesis~\cite{avidan1997novel}, 
3D shape representation~\cite{park2019deepsdf}, simulation, path planning~\cite{raja2012optimal}, 
3D modeling~\cite{tang20203d}, and image approximations.
We first performed an algorithmic analysis of the \nG
applications and found that all the \nG applications require 
a \MLP to learn the scene representations and an input encoding kernel to 
capture high frequency information in the visual data. 
We studied a wide range of input encoding algorithms and 
carefully picked three representative input encoding algorithms 
for further study, 1) {\em \ieOne encoding}, 2) {\em \ieTwo encoding} and 3) {\em \ieThree encoding}.
%\textcolor{red}{
We analyzed the above four \nG applications for all three encoding types,
on a modern desktop class GPU (RTX 3090). 
Our profiling shows that, 
%there is the gap of
%$\sim 2OOM-4OOM$ between
% typesetting to write OOM 
%the desired performance and power and the state of the art.
%We studied four representative \nG applications and identified the performance and power gap between the 
%desired targets and the state of the art.
%E.g. 
if we want to render 4k resolution 
frames at 60 FPS, there is a gap of 
$\sim\gapfpsVHGF \times$ to $\gapfpsNHGF \times$
between
the desired performance and the state of the art.
%For AR and VR applications, 
%there is an even larger gap of
%$\sim$ 2-4 OOM in the desired performance for the required power.
%}
This motivates the need to provide hardware acceleration to 
bridge this gap.

\begin{figure}
    \centering
    \includegraphics[width=\linewidth]{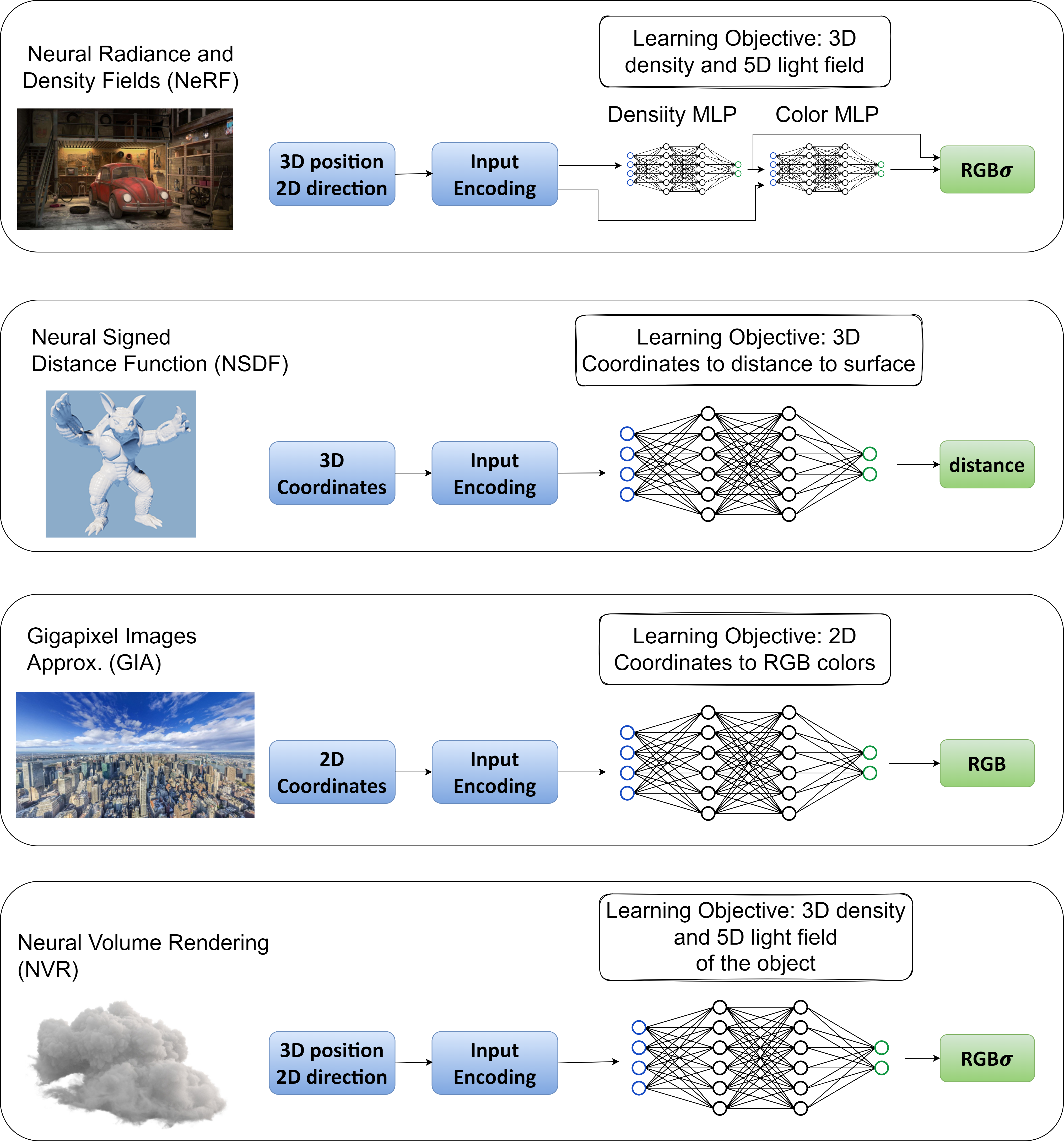}
    \caption{\small Structure of the four neural graphics applications under study.}
    \label{fig:all_app_struct}
        \vspace{-0.2in}
\end{figure}

In order to understand the performance bottlenecks of the representative 
\nG applications, 
we %used Nsight Compute~\cite{nsight} to 
perform 
kernel level performance breakdown analysis.
Our results show that input encoding and \MLP are the two 
most expensive kernels in all \nG applications
consuming
%On average across all \nG applications,
$\ieOneIEMLP \%$, 
$\ieTwoIEMLP \%$ and
$\ieThreeIEMLP \%$
of application time 
%is consumed by the 
%\IE and the \MLP kernels, 
for 
{\em \ieOne encoding}, 
{\em \ieTwo encoding} and
{\em \ieThree encoding} respectively.
Based on these results,
%Since our profiling motivates the need to accelerate 
%the \IE and \MLP kernels, 
we design an architecture 
 - a {\em  \NFP } (Figure~\ref{fig:arch}) - to accelerate 
these stages in hardware using an \IE engine and a hardware MLP engine. The \IE engine directly supports operations and dataflow identified by 
%Our hardware architecture is motivated by 
the kernel-level analysis of the 
\IE (Section~\ref{sec:understandingPerf}).
The MLP engine is similarly optimized for the small MLPs common in \nG \MLP kernels.
Moreover, in neural graphics, the outputs of the \IE kernel
are always consumed by the \MLP kernel (Figure ~\ref{fig:all_app_struct}) 
We exploit this fact in the \NFP 
hardware by fusing the \IE and \MLP engines. 

We propose a scalable \NPC (\npc{}) - consisting of several \NFP units -  along with 
the existing GPC units. 
%We also combined the input pre-processing kernels as one fused 
%kernel and the output post processing kernels as another fused kernel,
%in Vulkan~\cite{vulkan}, which leads to
%$\sim\fuserest\times$ kernel level performance improvement compared to Nvidia's 
%"un-fused" implementation~\cite{muller2022instant} of the pre-processing 
%and the post-processing kernels. 
We evaluate the performance of the 
\nG applications on our proposed architecture
for the scaling factor (number of \NFP units) of 8, 16, 32 and 64. 
For {\em \ieOne encoding},
%on average across the four \nG applications, 
the performance 
benefits of our architecture are
$\npcOAvgHG\times$,
$\npcTAvgHG\times$, $\npcFAvgHG\times$ and 
$\npcEAvgHG\times$ 
for the scaling factor of 8, 16, 32 and 64
respectively.
Our results show that with \ieOne encoding, 
\npc{} enables the rendering of 
4k Ultra HD resolution frames at 30 FPS for NeRF and
8k Ultra HD resolution frames at 120 FPS for all our 
other neural graphics applications. 
%30 FPS HD frames for NeRF,
%20 FPS QHD (2k resolution) frames
%and 30 FPS 5k resolution frames for NSDF,
%60 FPS 5k resolution,
%90 FPS Ultra HD (4k resolution) 
%and 
%120 FPS QHD (2k resolution)
%frames for NVR,
%and 120 FPS 5k resolution
%frames for GIA.

Our work makes the following contributions:
\begin{itemize}
    \item Neural graphics applications require dedicated hardware acceleration for real-time rendering. We 
    quantify the performance gap between the modern hardware and the desired performance targets.
 %   \textcolor{red}{
 %   Our profiling shows that, 
    %there is the gap of $\sim 2-4 OOM$ between the desired 
    %performance and power and the state of the art.
  %  if we want to render 4k resolution
  %  frames at 90FPS there is the gap of 
    $\sim\gapfpsVHGF \times$ to $\gapfpsNHGF \times$ 
    for rendering 4k resolution frames at 60 FPS.
  %  between
  %  the desired performance and the state of the art.
  %  For AR and VR applications, 
  %  there is the gap of
   % $\sim$ 2OOM-4OOM 
   % between
   % the desired performance and power and the state of
   % the art.}
    
    \item We studied the \nG applications and identified that the \IE and the \MLP kernels are common
    performance bottlenecks. 
    \iffalse
    Averaged across all four representative \nG applications,
    $\ieOneIEMLP \%$, $\ieTwoIEMLP \%$ and $\ieThreeIEMLP \%$ of application time 
    is consumed by the \IE and the \MLP kernels, 
    for {\em \ieOne encoding}, {\em \ieTwo encoding} and {\em \ieThree encoding}, respectively. 
\fi
   
    \item We present an efficient hardware architecture to support \nG applications in real time. Our architecture  accelerates the \IE and \MLP kernels through dedicated engines and fuses the engines for a more efficient dataflow. We show that our hardware 
    architecture is scalable and flexible enough to support a wide range of \nG applications. 
    
    \item We quantify the benefits of our hardware architecture and show 
    significant performance benefits against GPU baseline for all four \nG applications and 
    three \IE types. 
\iffalse
    
    For {\em \ieOne encoding} on average across the four \nG applications, 
    the performance benefits of our architecture are
    $\npcOAvgHG\times$, $\npcTAvgHG\times$, $\npcFAvgHG\times$ and 
    $\npcEAvgHG\times$ 
    for the scaling factor of 1, 2, 4 and 8
    respectively.
    \textcolor{red}{
Our results show that with \ieOne encoding, 
for NeRF application 
\npc{} enables the rendering of 30 FPS HD frames,
for NSDF application
\npc{} enables the rendering of 120 FPS QHD (2k resolution) frames
and 30 FPS 5k resolution frames,
for NVR application
\npc{} enables the rendering of 
60 FPS 5k resolution,
90 FPS Ultra HD (4k resolution) 
and 
120 FPS QHD (2k resolution)
frames,
for GIA application
\npc{} enables the rendering of  
120 FPS 5k resolution
frames.
}
\fi
\end{itemize}

\section{Neural Graphics: An Overview}
\label{sec:background}

\begin{figure}
    \centering
    \includegraphics[width=\linewidth]{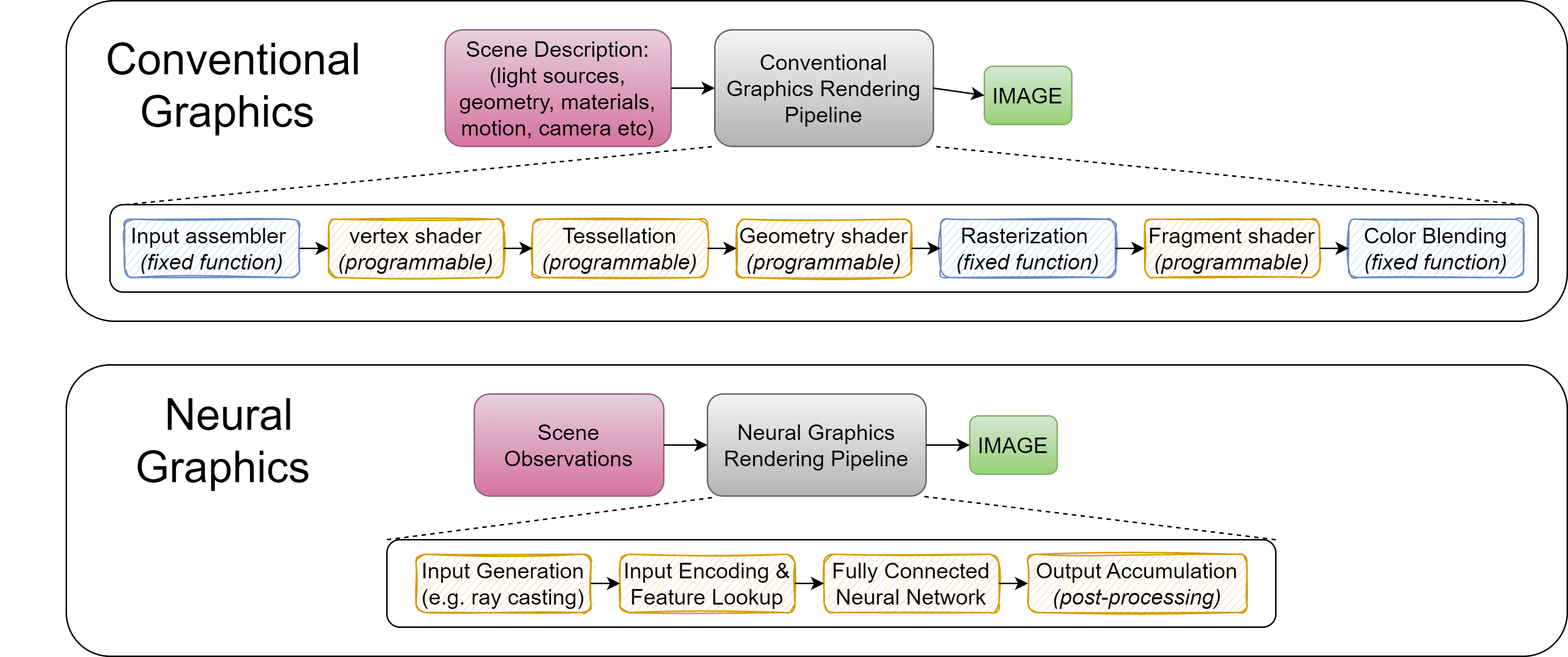}
    \caption{%\textcolor{red}
    {\small Rendering pipeline: Conventional Computer Graphics~\cite{vulkan} VS Neural Graphics~\cite{tewari2022advances}.}
    }
    \label{fig:cgvsng}
    \vspace{-0.2in}
\end{figure}

Figure~\ref{fig:cgvsng} depicts a high-level 
comparison of conventional  
rendering pipeline and 
the neural graphics pipeline.
Rendering in conventional real-time graphics starts with 
the detailed description of the physical and material properties 
of the scene.
This description is passed as input to the rendering 
pipeline which then generates a 2D image (frame) as output.
In neural rendering, the description of the 
geometric and material properties of the scene are derived from multiple scene observations (images or video), which serves as an 
input to the neural rendering pipeline. 
Neural graphics replaces the conventional 
rendering pipeline with much simpler 
neural rendering pipeline.
The neural rendering pipeline at high level consists 
of three stages.
\iffalse
1) {\em Pre-processing stage} which is 
responsible for generating the input features for the 
neural networks from the given positional coordinates.
2) {\em Inference stage} that  
computes the forward pass of the fully-connected neural 
networks 
and outputs the 
color and density values corresponding to the input features.
3) {\em Compositing stage} that blends 
the color based on density values to generate the final 
pixels to be displayed. 
Section~\ref{sec:understandingPerf} provides further 
details about these stages. \\

As discussed above, the objective of a typical neural graphics application is to be able to generate photo-realistic imagery 
of a scene (3D or 2D) by learning the detailed visual information about the scene in a neural network. Below we describe how a neural rendering pipeline is implemented to render photo-realistic visual data.

\subsection{Neural Rendering Pipeline}
Figure~\ref{fig:ng_app_struct}-a shows the structure of a typical neural graphics pipeline.
A typical neural graphics pipeline consists of three stages \\
\fi

{\bf 1) Input stage:} This stage is responsible for generating the inputs to a neural network.  
In classical computer vision and image processing applications, the typical goal is image classification 
or image transformation. In these applications,  
an image is typically used as input to the 
neural network. The input layer of the network can be vary large depending upon the 
resolution of the image. For example, if a network is processing a $224\times224$ RGB image (as in case of AlexNet~\cite{alom2018history}),
the input layer has $224\times224\times3$ dimensions. %The scales up with the resolution of the image.
Unlike classical computer vision and image processing applications, where usually an image is 
passed as input to the neural network, in \nG applications, the
neural network inputs are either encoded
pixel positions/coordinates $(x,y,z)$ or encoded pixel positions/coordinates and camera viewing angle $(x,y,z,\theta,\phi)$,
depending upon the application under consideration. \\

{\bf 2) Inference stage:} This stage is responsible for learning the detailed information 
about the scene. 
The learning objective of the \nG applications is different from a typical 
computer vision and image processing application. 
In a typical computer vision and image processing application,  
the goal is usually to learn common features from a batch of training images and then, 
during inference phase, classify the images into a set of classes by identifying the features in them.
The fully connected layers on the input side of a neural network usually become very expensive 
for a typical computer vision application and hence the neural networks used 
in classical computer vision applications are usually 
convolutional neural networks.
%The features of the input images are learned in small filters of convolutional layers; 
%typically a filter size of $3\times3$ or $5\times5$ is used. These 
%networks have convolutional layers in the beginning stages of the network, where filters are used 
%to reduce the feature dimensions,
%and a few fully connected layers in the later half of the network, where the dimensions of the 
%intermediate features are relatively smaller.
%A fully connected layer has more parameters than a convolutional layer and hence the 
%learning capacity of a fully connected layer is much more than a convolutional layer.
In a \nG application, 
%the goal is to learn the detailed representation of the scene, 
%which is usually very information dense visual data.
%Hence, due to the information dense nature of the learning data, fully connected 
%networks, as opposed to convolutional networks, are used to learn the scene representation in \nG applications. 
%\textcolor{red}{
the goal is to learn the detailed representation of the scene, which usually comprises high-frequency visual data. Suitable input encodings such as frequency encoding~\cite{mildenhall2021nerf} and parametric grid encodings~\cite{muller2022instant} combined with strong compression capabilities of fully connected networks have been shown to learn neural representations for such information well.
%}

In training phase, the scene observations are learned by the neural network using classical 
neural network training techniques (for example gradient descent and Adam optimization),
where the loss function propagates the gradients in the 
backward direction and adjusts the weights of the neural network. 
In the inference phase, the network outputs the 
pixel color $(RGB)$ or pixel color and density information $(RGB,\sigma)$ using the given
pixel coordinates/positions $(x,y,z)$ or pixel coordinates/positions and camera viewing angle $(x,y,z,\theta,\phi)$
as inputs. \\

{\bf 3) Compositing stage:} The output of the fully-connected neural network is either 
three channel pixel color $(RGB)$ or pixel color and density information $(RGB,\sigma)$. 
The goal of the output stage is to accumulate the color and density information of the 
individual pixels and assemble the final output imagery. 
Classical volume rendering techniques~\cite{drebin1988volume, lichtenbelt1998introduction, westover1989interactive} can be used to project the output colors and densities into an image. \\

\begin{figure}
    \centering
    \includegraphics[width=\linewidth]{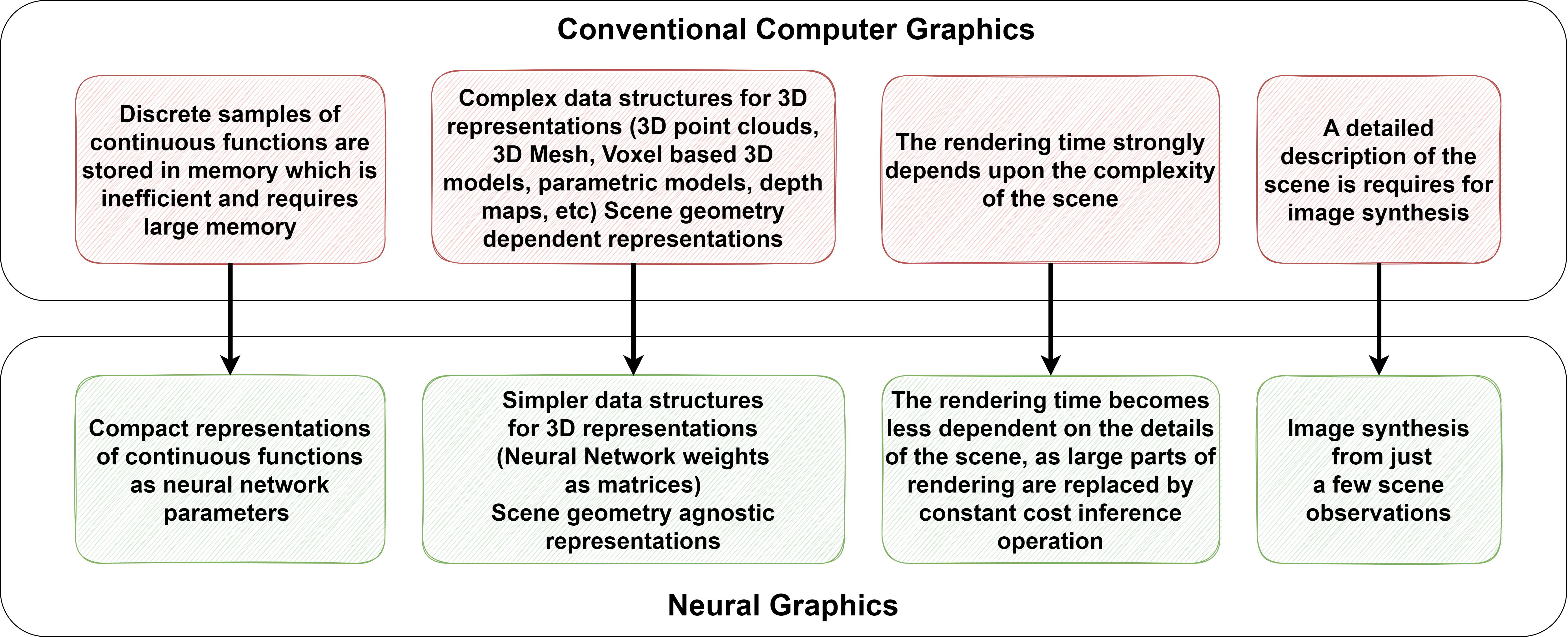}
    \caption{%\textcolor{red}
    {\small Inherent benefits of Neural Graphics over 
    Conventional Computer Graphics.}}
    \label{fig:cgvsngbenefits}
        \vspace{-0.2in}
\end{figure}

Figure~\ref{fig:cgvsngbenefits} shows the 
high level benefits of neural graphics over conventional
computer graphics. 
1) {\bf More compact representations of scenes:}
In classical computer graphics, "Fields" are widely 
used to 
parameterize the physical properties of an
object or scene over space and time.
Such space-time parameterizations, defined for all spatial 
and/or temporal coordinates, are  
needed to synthesize 3D
shapes and/or 2D images.
%Continuous functions~\cite{van2003graphsplatting} are typically used to faithfully
%represent a field.
%These continuous functions are then 
%sampled and stored as discrete samples
%in computers; 
%the sampling rate must be at least the Nyquist sampling rate
%to avoid aliasing. 
%As the complexity of the scene 
%grows, the memory requirement to store these samples 
%explodes. 
%\textcolor{red}{
In order to faithfully store arbitrary functions by way of classic discrete samples, high sampling rates are needed to avoid aliasing. As the complexity of the scene grows, the memory requirement to store these samples explodes.
%}
However, 
in neural graphics, the field is parameterized, 
fully or in part,
by a neural network. Such parameterized fields are known as 
neural fields.
%The parameters of neural networks are inherently 
%continuous and do not require Nyquist sampling for storage.
%\textcolor{red}{
Neural representations are known to adapt well to continuous functions with sparse discontinuities even with much lower storage capacities.
%}
Hence, neural fields enable compact and efficient 
representations of the scene.
2) {\bf Simpler domain-agnostic data structures:} 
In conventional computer graphics, 
complex data structures such as 3D point clouds~\cite{guo2020deep}, 3D meshes~\cite{wang2018pixel2mesh},
voxel based 3D models~\cite{tahir2021voxel},
parametric models,
and depth maps~\cite{reeves1987rendering} are used  
for 3D representations. 
These data structures are scene geometry dependent.
%Whereas, in neural graphics, the 3D scenes are learned 
%in the weights of the neural networks.
%These weights are stored as matrices which are 
%much simpler data structures compared to 
%conventional 3D representations.
%Moreover, unlike conventional 3D representations, 
%the weight matrices of a neural network are scene geometry 
%agnostic. 
%\textcolor{red}{
%While, 
In neural graphics, while aspects of such data structures may still be required to limit network complexity at larger scales, extended neural representations often lead to more unified, higher-level data structures and thus reduced complexity.
%}
3) {\bf Predictable performance:}
%In conventional computer graphics, the rendering time is 
%non-deterministic as it 
%depends upon the complexity of the scene.
%In neural graphics, as "rendering" is effectively 
%replaced by the inference operation of the neural networks, the
%rendering time is deterministic as the inference time 
%of neural networks is 
%inherently
%deterministic.
%\textcolor{red}{
In conventional computer graphics, the rendering time strongly depends on the complexity of the scene. In neural graphics, as large parts of the rendering are replaced by a constant-cost inference operation, the rendering time becomes less dependent on details of the scene, and instead is mostly a function of scale and higher-level scene layout.
%}
4) {\bf Higher-level scene definition:}
As explained earlier, in conventional computer graphics 
a complete description of the physical and material properties of
the scene is required for rendering. In neural graphics, however, 
the properties of the scene can implicitly be learned from 
a few images or video of the scene. This leads to much simpler scene definitions.
%why? \\

The above benefits make neural graphics arguably the biggest advance in the field of computer graphics in decades.

%\begin{itemize}
%    \item what is it? 
%    \item why
%    \item types (applications)
%    \item F-a CG vs NG
%    \item F-b NG in systems benefits
%\end{itemize}

\begin{figure*}
    \centering
    \includegraphics[width=\linewidth]{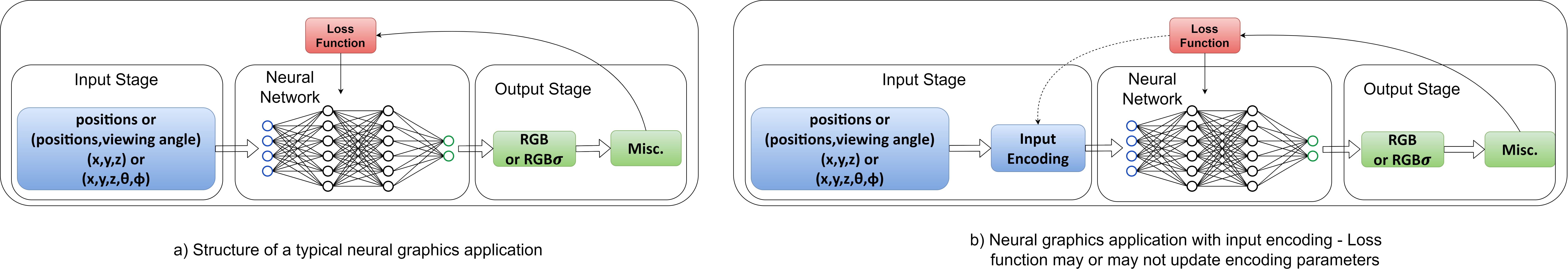}
    \caption{\small Different stages of a neural graphics pipeline.}
    \label{fig:ng_app_struct}
        \vspace{-0.2in}
\end{figure*}

\subsection{Input Encoding}
Photo-realistic visual data usually has high frequency information. 
For example, the crisp RGB colors of a gigapixel image, 
the detailed texture, lighting, and geometry information of a 3D scene, etc., 
are some of the examples of high frequency visual data. 
In \nG, \MLPs are used to learn and represent this high 
frequency visual data.
Previous works have shown that \MLPs (Figure~\ref{fig:ng_app_struct}-a) are biased towards learning low frequency information~\cite{rahaman2019spectral, mildenhall2021nerf}
of the given data and are not good function approximators when high frequency information 
needs to be captured. 
In order to solve the problem of learning high frequency visual information using 
multi-layer perceptron, 
%\textcolor{red}{
the original NeRF paper~\cite{mildenhall2021nerf} 
(Vanilla-NeRF)
introduced the idea of input encoding.
%} 
The idea of input encoding is to map the low dimensional input vectors to higher dimensional space 
using a mapping function (Figure~\ref{fig:ng_app_struct}-b).
The mapping function can either be a fixed high-frequency function ($\sin,\cos,$ fourier~transform)
or a learn-able function (embeddings, neural networks). 
Based on the mapping function used, the input encoding schemes can be divided into two high level categories: 
{\em 1) fixed-function encodings 2) parametric encodings}. \\

{\em \bf 1) Fixed-function encodings:}

Vanilla-NeRF~\cite{mildenhall2021nerf} was the first paper to show that, mapping the 5D vector of input positions and 
viewing angle to a higher dimensional space, by using high frequency $\sin$ and $\cos$ 
functions, and using the outputs of the high frequency functions as 
input features to the multi-layer perceptron, enables the \MLP to learn the high frequency variations of the scene.
Since vanilla-NeRF, there had been a lot of work on improving the input encoding. 
%\textcolor{red}{
Tancik et al.~\cite{tancik2020fourier} showed, for example, that replacing $\sin$ and $\cos$ with  
Fourier transform 
lets the network learn
high frequency functions in low dimensional domains.
%}
These encoding schemes are known as {\em frequency encoding schemes}.
as they use high frequency sinosoids 
for mapping the low dimensional inputs to higher dimension space.
%~\cite{muller19}~\cite{muller20} has shown that one blob encoding, which is essentially a continuous variant of the
%one-hot encoding~\cite{onehot} can achieve more accurate results than frequency encoding schemes in bounded domains.
%But their proposed encoding is not scale-able which makes it infeasible for unbounded scenes.
All these schemes are fixed function encoding schemes as they use a fixed compute function 
for mapping the low dimensional input vector to higher dimensional space. \\

{\em \bf 2) Parametric encodings:}
Recently, state of the art results have been shown by {\em parametric encodings}. Parametric encoding schemes
essentially use additional trainable parameters in addition to weights and biases of the 
neural networks to learn information about the scene.
These parameters can be arranged as auxiliary data structures 
and used to map low dimensional positions to 
higher dimensional inputs, which are then used to query the neural network.
\iffalse
There has been a lot of research on the types of data structure to be
used for parametric encoding. 
~\cite{chaber20}~\cite{jiang20}~\cite{liu20}~\cite{mehta21}~\cite{pend20a}~\cite{sun21} and ~\cite{yu21a} 
used grid as auxiliary data structures to arrange and look-up additional parameters for neural network inputs  
whereas ~\cite{takikawa21} used trees. 
Although, the parametric encoding schemes yield strictly better fidelity in the output imagery than 
non-parametric encoding schemes~\cite{}, they usually 
result in complex training routines and the explicit 
data structures used to arrange/look-up the additional 
trainable parameters are not versatile.
~\cite{ngp} proposed that instead of using explicit auxiliary data structures (Grids and Trees)
\fi
In a popular class of parametric schemes~\cite{muller2022instant, muller2021real}, the trainable parameters are stored in generic 
lookup tables where the number of parameters are 
picked based on the desired reconstruction quality. 
These works  propose to divide the scene into multiple resolution levels (grid) and 
use a separate lookup table for each resolution level.
As the scene is divided into multiple grids, the approach is called 
{\em multi-resolution grid encoding} and the number of 
lookup tables (or resolution levels) can also be optimized as a hyper parameter.
Different resolution levels give different output fidelity, ~\cite{muller2022instant} has 
shown that for most \nG applications, depending upon the type of grid used, 
8 to 16 resolution levels produce 
acceptable visual fidelity. \\

\begin{figure}
    \centering
    \includegraphics[width=\linewidth]{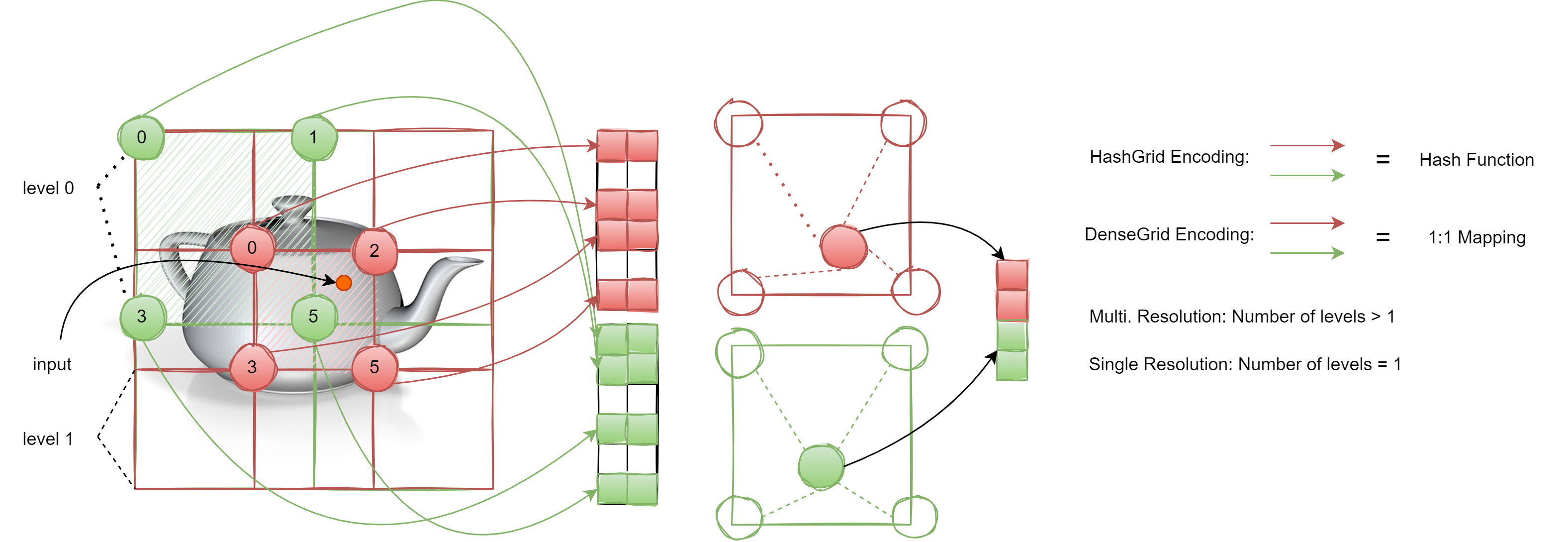}
    \caption{Illustration of {\em Multi resolution grid encoding}.}
    \label{fig:all_encs}
    %\vspace{-0.2in}
\end{figure}

The different steps performed in {\em multi-resolution grid encoding} are 
explained in Figure~\ref{fig:all_encs}.
The encoding parameters are arranged into L levels, 
each containing up to T feature vectors with dimensionality F.
Each level is independent and stores 
the feature vectors at the vertices of a grid. 
Each corner of the grid is mapped to 
an entry in the level’s respective feature vector array where each 
feature vector array has a fixed size of T. 
The mapping between the grid and the feature vector array is 1:1 for 
coarse levels because  the dense grid
requires fewer than T parameters for coarse levels. 
%\textcolor{red}{
However, for finer levels, the feature vector can either be mapped 1:1 or 
treated as hash table and 
a hash function can be used to index into the 
array~\cite{muller2022instant}.
%}
The input encoding, where a hash function is used to index into lookup table, is 
called {\em multi-resolution hashgrid encoding}. 
Whereas, the encoding where 1:1 mapping is used for all the resolution levels 
is known as {\em multi-resolution densegrid encoding}. 
The hash function used by the state of the art parametric encoding scheme (instant-NGP~\cite{muller2022instant})
is presented in Equation~\ref{eq:hash}.
\begin{equation}
    \label{eq:hash}
    h(x) = (\oplus_{i=1}^{d} x_i \pi_i)~mod~T
\end{equation}
where $\oplus$ denotes the bit-wise XOR operation and $\pi$
are unique large prime numbers. 
The feature vectors at each corner of the grid 
are linearly interpolated to ensure continuous 
representation. 
The interpolated feature vectors for all the levels 
are then concatenated to generate the final encoded input to the 
multi-layer perceptron, as shown in Figure~\ref{fig:all_encs}. 
%Instant-ngp does not use
%any explicit collision handling mechanism;
%potential hash collisions are resolved implicitly by 
%relying on training (gradient-based optimization) to store the 
%appropriate sparse details in the array and then using 
%neural network to implicitly approximate the collision resolution.
The number of training parameters are bounded by $T\times L\times F$. 
~\cite{muller2022instant} has shown that for {\em multi-resolution hashgrid encoding} only using 16 levels and 2 features per 
array entry produces high fidelity image frames for many 
neural graphics applications, hence the number of 
trainable encoding parameters is set $T\times 16\times 2$ where 
T ranges from $2^{14}~to~2^{24}$ depending upon the neural graphics application and  
the desired image fidelity.
For {\em multi-resolution densegrid encoding} 2-8 levels can be used to get high visual fidelity 
outputs. \\

\section{Does Neural Graphics Need Hardware Support?}
To understand the performance characteristics of neural graphics, we focus on the following four 
representative 
neural graphics applications 
%\textcolor{red}{
that cover a relatively wide range of graphics tasks, as presented by Müller et al.~\cite{muller2021real},
%}
including rendering, 
novel view synthesis, 3D shape representation, simulation, path planning, 3D modeling and image approximations:
%We will also be using these following four applications as benchmarks. 

1) {\bf Neural radiance and density fields (NeRF):} 
The learning objective during training phase is 
is to learn the 3D density and the 5D light 
field of a given scene from a few scene observations (images or video of the scene).
The structure of the NeRF application is shown in Figure~\ref{fig:all_app_struct}.
In NeRF, two fully-connected neural networks, 1) Density MLP and 2) Color MLP, are concatenated together 
where density MLP learns the density information $(\sigma)$ and 
the color MLP learns the view dependent $(RGB)$ color information of the scene. 
As the density information of a scene is view direction agnostic, 
the encoded coordinates/positions $(x,y,z)$ are used as input to the density MLP. 
As the color information depends upon position as well as view angle, 
the output of the density MLP along with the encoded view directions $(\theta,\phi)$ are 
passed as input to the color MLP. 
%\textcolor{red}{
The output of the color MLP is a 
three dimensional 
vector containing 
the pixel color information $(RGB)$. 
The density information is then concatenated with the 
pixel color information to get the 
four dimensional vector containing the 
pixel color and the density
information $(RGB,\sigma)$.
%} 
\\

2) {\bf Neural signed distance functions (NSDF):} 
In classical computer graphics, signed distance functions (SDFs) are used to represent a 3D shape 
as the zero level-set of a function of position x.
In graphics, SDFs are commonly used in applications such as simulation, path planning,
3D modeling, and video games~\cite{muller2022instant}. In neural approximations of SDFs, the MLP learns the
mapping from 3D coordinates to the distance to a surface.
The structure of the NSDF application is shown in Figure~\ref{fig:all_app_struct}.
The inputs to the MLP are encoded positions 
and the final output is 
the distance to the surface. \\

3) {\bf Gigapixel image approximation (GIA):} 
A gigapixel image is an ultra high definition digital image bitmap, 
usually made by combining multiple detailed images into a single image.
A Gigapixel image has billions of pixels, which is much more than the capacity of a normal professional camera.
In GIA a neural network is used to approximately learn the 
gigapixel image in its trainable parameters. 
The learning objective of the MLP in GIA application is to
learn the mapping from 2D coordinates to RGB colors of the gigapixel image. 
The structure of the GIA application is shown in Figure~\ref{fig:all_app_struct}.
As the MLP learns the 2D image, there is no density information and hence
the view direction is also not required; 
the inputs to the MLP are the encoded pixel positions and the outputs are 
the corresponding pixel color $(RGB)$.
This application can be considered as an 
important benchmark to test neural network's ability to learn the high frequency details of visual data. \\

4) {\bf Neural volume renderer (NVR):}
This application is similar to NeRF. 
%the only difference being that 
%instead of an unbounded 3D scene, the network learns the 
%the 3D density and the 5D light 
%field of a bounded 3D object.
%\textcolor{red}{
The only difference is that, 
instead of learning the density and the emission field, 
the network in neural volume rendering learns the
density and a reflectance field, which can be used to simulate the light transport in the volume using path tracing.
%with proper relighting etc.
%}
The structure of the NVR application is shown in Figure~\ref{fig:all_app_struct}.
In NVR, two fully-connected neural networks, 1) Density MLP and 2) Color MLP, 
are concatenated together 
where the density MLP learns the density information $(\sigma)$ and 
the color MLP learns the view dependent $(RGB)$ color information of the bounded object. 
Similar to NeRF, the encoded positions $(x,y,z)$ are used as input to the density MLP and
the output of the density MLP along with the encoded view directions $(\theta,\phi)$ are 
passed as input to the color MLP.
The output is a four 
dimensional vector containing 
the pixel color and the density information $(RGB,\sigma)$. \\

Each application can be implemented using a variety of input encodings. 
\iffalse
We focus on parametric encodings, instead of fixed-function encodings. 
In parametric encoding schemes, as learnable parameters are embedded into the input of \MLPs (input embedding),
some of the learning task is offloaded to the input embedding.
This allows the use of smaller neural networks instead of huge networks used along with frequency encoding schemes. 
Existing work~\cite{ngp}~\cite{nrc} has shown that {\em multi-resolution grid encodings} 
along with tiny MLPs (2-4 hidden layers and only 64 hidden neurons per layer),
can learn high frequency 
data for a variety of \nG applications.
\fi
As parametric encodings produce strictly better output fidelity than frequency encodings~\cite{muller2022instant, muller2021real, chabra2020deep, jiang2020local},
we picked parametric encoding for further exploration.
In order to faithfully represent the state of the art in parametric encodings, 
we explored three different types of {\em parametric encodings} in this work --
{\em 1) Multi resolution hashgrid encoding:} Hash function is used to generate the indices of the lookup tables while 
mapping the grid features to the lookup tables entries. The number of resolution levels used is 16~\cite{muller2022instant, ngpgithub}.
{\em 2) Multi resolution densegrid encoding:} 1:1 mapping is used between grid features and the lookup tables entries. 
The number of resolution levels used is 8~\cite{muller2022instant, ngpgithub}
{\em 3) Low resolution densegrid encoding:} 1:1 mapping is used between grid features and the lookup tables entries. 
The number of resolution levels used is 2~\cite{muller2022instant, ngpgithub}

We profiled the above four \nG applications
with the chosen three input encoding schemes using the open source code 
published by 
%Nvidia~\cite{ngpgithub}.
%\textcolor{red}{
Müller et al~\cite{ngpgithub}.
%}
The parameters for all our \nG applications and encoding schemes are shown in 
Table~\ref{tab:app_struct}.
%For evaluation,
%In order to evaluate our applications on GPU, 
%we  The applications are 
%implemented in CUDA, where 
Both the input encoding algorithm and the \MLP are implemented as separate fused CUDA kernels~\cite{muller2021real, ngpgithub, muller2022instant}. Unlike standard MLPs the fully-fused MLPs 
do not have any explicit biases. 
Due to the smaller size of the 
fully-fused MLPs, the intermediate activations and the partial sums
can be stored on the faster on-chip memory, reducing the number 
of global memory accesses. 
%Nvidia's implementation of \nG MLPs as one fully-fused CUDA kernel makes sure that
%all the intermediate features stay on the fast on-chip memory of the 
%GPU and the only global memory accesses are for loading weights and the input 
%activations, and storing back the output activations. 
A python based wrapper generates the inputs and displays the final rendered frame. 
We run the applications on Nvidia's RTX3090 using CUDA version 11.7 and report the total runtime of the applications. 

\begin{figure*}[ht!]
   \subfloat[\label{genworkflow}]{%
      \includegraphics[width=0.32\textwidth]{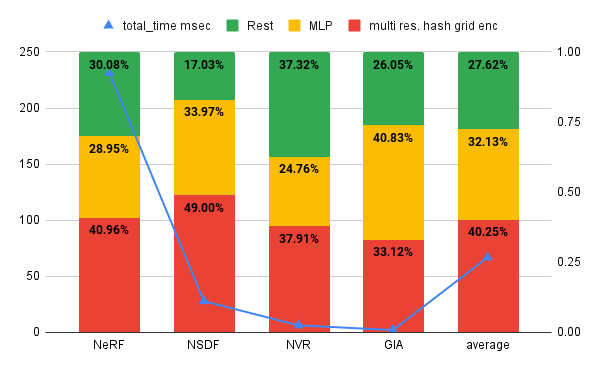}}
\hspace{\fill}
   \subfloat[\label{pyramidprocess} ]{%
      \includegraphics[width=0.32\textwidth]{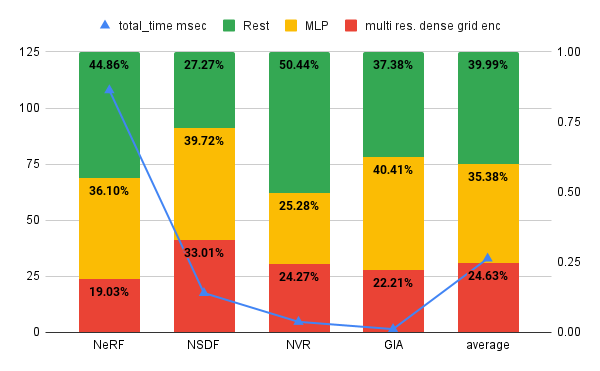}}
\hspace{\fill}
   \subfloat[\label{mt-simtask}]{%
      \includegraphics[width=0.32\textwidth]{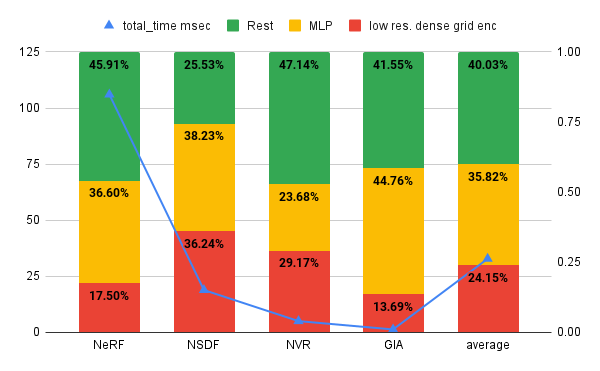}}\\
\caption{\small Kernel level performance breakdown analysis of \nG applications: The number of cycles consumed by the 
different kernels of the \nG applications (as percentage 
of total application cycles). 
(a) {\em \iEOne encoding},  
(b) {\em \iETwo encoding},
(c) {\em \iEThree encoding.}}\
\label{fig:app_brkdn}
    %\vspace{-0.2in}
\end{figure*}

Our results (Figure~\ref{fig:app_brkdn}) show that, 
on a modern desktop class GPU (RTX3090),
for \ieOne encoding,
rendering $\sim2M$ pixels ($1920\times1080$ frame) 
takes 
$\timeNHG msec$, $\timeSHG msec$, $\timeIHG msec$ and $\timeVHG msec$
for NeRF, NSDF, GIA and NVR respectively.  
This is unacceptable for real-time applications. 
For instance,
if we want to render 4k frames at 60FPS 
only one of our \nG applications (GIA)
is able to meet that target, 
there is the 
performance gap of 
$\gapfpsNHGF \times$, $\gapfpsSHGF \times$, 
and $\gapfpsVHGF \times$
for NeRF, NSDF and NVR respectively.

In order to understand the performance bottlenecks of \nG applications, 
we %used Nsight Compute~\cite{nsight} to 
perform 
kernel level performance breakdown analysis.
Figure~\ref{fig:app_brkdn}
%, we present the number of cycles consumed by the 
%different kernels of the \nG applications for different input encoding types, as percentage of 
%total number of cycles spent consumed by the entire application.
%Our results 
shows that input encoding and \MLP are the two 
most expensive kernels in all \nG applications. 
For {\em \ieOne encoding}, on average across all \nG applications, $\ieOneIEPercent \%$ of the total number of cycles are 
consumed by the input encoding kernel 
and $\ieOneMLPPercent \%$ cycles are consumed by the \MLP kernel. 
On average, this amounts to $\ieOneIEMLP \%$ of application time.  
For {\em \ieTwo encoding}, on average across all \nG applications, $\ieTwoIEPercent \%$ of the total number of cycles are 
consumed by the input encoding kernel 
and $\ieTwoMLPPercent \%$ cycles are consumed by the \MLP kernel. 
On average, this amounts to $\ieTwoIEMLP \%$ of application time.  
For {\em \ieThree encoding}, on average across all \nG applications, $\ieThreeIEPercent \%$ of the total number of cycles are 
consumed by the input encoding kernel 
and $\ieThreeMLPPercent \%$ cycles are consumed by the \MLP kernel. 
On average, this amounts to $\ieThreeIEMLP \%$ of application time.  
This data motivates the need to accelerate the 
input encoding and \MLP kernels. 
In next section, we will provide a further breakdown of the time spent in the input encoding and the \MLP kernels to understand their 
performance bottlenecks.

% \begin{itemize}
%    \item structure of a NG app (IE + MLP + Misc)
%    \item F-a structure
%    \item example apps 
%    \item F-b (one for each app, structure + function (opt))
%    \item example input encodings - introduce both classes, select with reasoning, discuss the selected one 
%    \item F-c one for each input encoding - only ones you selected 
%    \item Performance breakdown w and w/o SW optimization 
%    \item F-d app x IE pie/bar charts 
%    \item talk about samples 
    
%\end{itemize}

\section{Understanding Performance of Neural Graphics Applications}
\label{sec:understandingPerf}

\begin{figure}
\centering
\includegraphics[width=\linewidth]{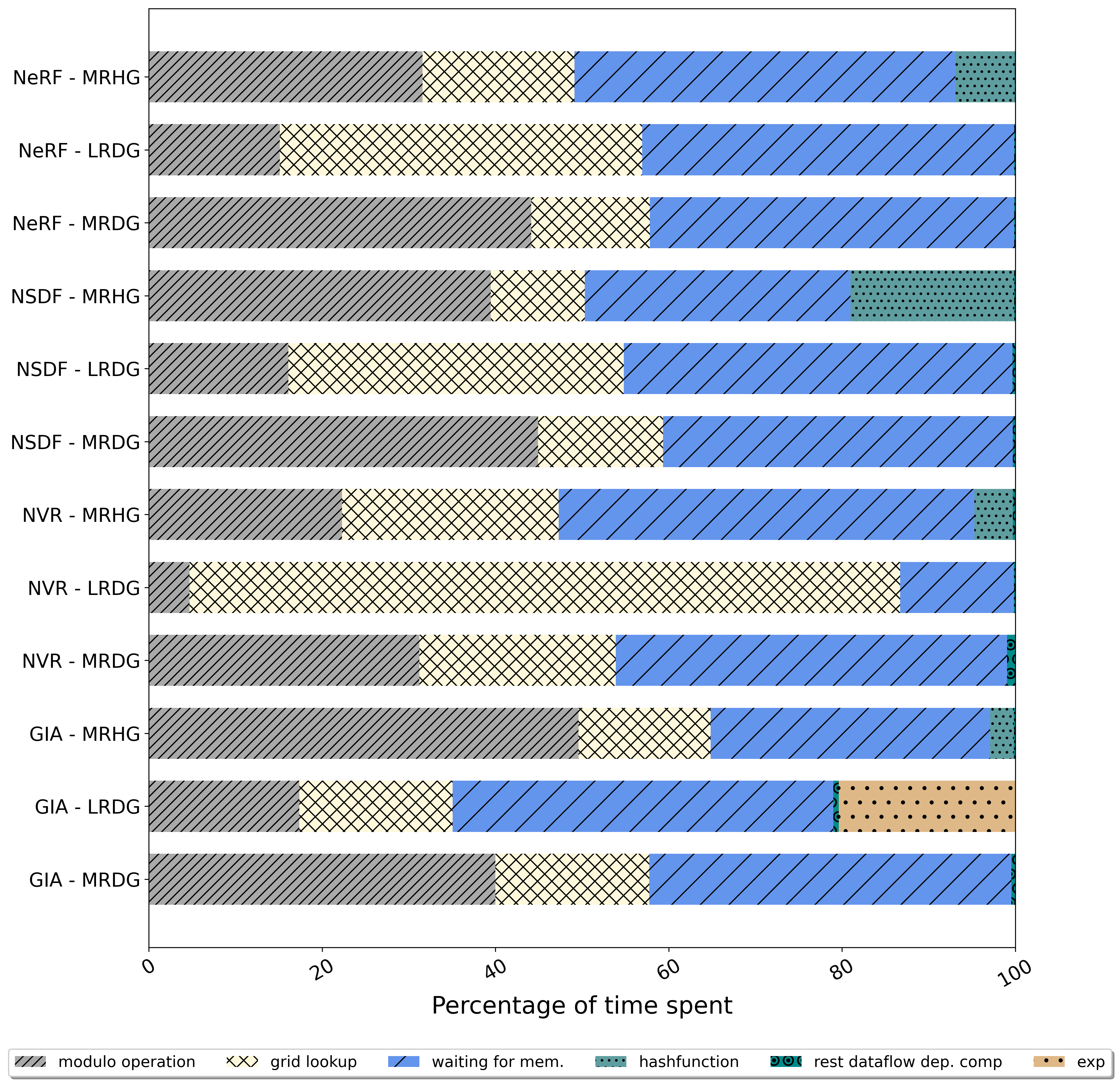}
% \begin{tabular}{ccc}
% \subfloat[\appN- \ieone]{\includegraphics[width =0.30\linewidth]{hashgrid_nerf_brkdn_sorted.png}} &
% \subfloat[\appN- \ietwo]{\includegraphics[width =0.30\linewidth]{densegrid_nerf_brkdn_sorted.png}} &
% \subfloat[\appN- \iethree]{\includegraphics[width =0.30\linewidth]{densegrid_1res_nerf_brkdn_sorted.png}}\\
% \subfloat[\appS- \ieone]{\includegraphics[width =0.30\linewidth]{hashgrid_sdf_brkdn_sorted.png}} &
% \subfloat[\appS- \ietwo]{\includegraphics[width =0.30\linewidth]{densegrid_sdf_brkdn_sorted.png}} &
% \subfloat[\appS- \iethree]{\includegraphics[width =0.30\linewidth]{densegrid_1res_sdf_brkdn_sorted.png}}\\
% \subfloat[\appV- \ieone]{\includegraphics[width =0.30\linewidth]{hashgrid_vol_brkdn_sorted.png}} &
% \subfloat[\appV- \ietwo]{\includegraphics[width =0.30\linewidth]{densegrid_vol_brkdn_sorted.png}} &
% \subfloat[\appV- \iethree]{\includegraphics[width =0.30\linewidth]{densegrid_1res_vol_brkdn_sorted.png}}\\
% \subfloat[\appI- \ieone]{\includegraphics[width =0.30\linewidth]{hashgrid_image_brkdn_sorted.png}} &
% \subfloat[\appI- \ietwo]{\includegraphics[width =0.30\linewidth]{densegrid_image_brkdn_sorted.png}} &
% \subfloat[\appI- \iethree]{\includegraphics[width =0.30\linewidth]{densegrid_1res_image_brkdn_sorted.png}}

% \end{tabular}
\caption{\small The operation level breakdown of the 
of input encoding kernels for different 
input encoding types. The 
five most expensive operations in terms of number of 
cycles spent.
\ieone: {\em \ieOne}, 
\ietwo: {\em \ieTwo}, 
\iethree: {\em \ieThree}.}
\label{fig:IE_brkdn}
    %\vspace{-0.2in}
\end{figure}

We performed further analysis (using Nvidia's nsight compute) to understand the performance bottlenecks in the \IE and \MLP kernels.
%and the 
%compute operations of the input encoding kernel, 
%we performed the operation level breakdown analysis 
%of the input encoding kernel for different 
%input encoding types using Nvidia's nsight compute.
Figure~\ref{fig:IE_brkdn} shows 
%the cycle level breakdown of the five most expensive 
%operations
%in an input encoding kernel for different encoding 
%types. 
%We used Nsight compute to perform
the operation level breakdown of the 
of input encoding kernels for different 
input encoding types.
%Figure~\ref{fig:IE_brkdn} shows the 
Five most expensive operations in terms of number of 
cycles spent are labeled in the 
Figure~\ref{fig:IE_brkdn}.
As explained in  Section~\ref{sec:background}, the grid lookups are the 
major building block of the input encoding algorithms; 
our analysis also shows that the grid lookups take significant amount of cycles across all three input encoding types. 

%Table~\ref{tab:util} 
The percentage utilization of the
GPU compute and memory resources,
%presents GPU resource utilization 
for the input encoding kernel
is presented in Table~\ref{tab:util}. On average, across all encoding kernel calls, 
the memory utilization of the GPU is higher than compute utilization. 
This is because input encoding is a memory intensive workload that requires 
performing lookup operations for mapping inputs to learned features 
%Moreover, t
even as the lookup tables for all the resolution levels do not entirely fit 
on the L2 cache of RTX3090. The memory wait time to resolve the cache misses 
also adds to the overall cycles. 
Our analysis also shows that 
the integer mapped modulo operation is one of the most expensive operations for all three input encoding types. 
%This is because the modulo operation gets mapped to the general purpose
%ALU of the GPU and cannot utilize the efficient tensor cores of the GPU.
%\textcolor{red}{
This is because the modulo operation gets mapped to the less efficient general integer modulo operation instead of using more efficient bitwise operations which account for the fact that hash map sizes are always powers of two.
%}

As \ieTwo and \ieThree have one-on-one mapping of grid indices, the hash function 
is not called for these input encoding types and hence the breakdown shows zero cycles for the hash 
function. However, the hash function consumes significant number of cycles for \ieOne. 
This is because the hash function 
%requires XOR operation which cannot 
%exploit the efficient tensor cores of the GPU and is instead scheduled on the 
%general purpose ALU of the GPU.
%\textcolor{red}{
cannot efficiently utilize the ALU due to the stalls caused by waiting on the long scoreboard to resolve 
the global memory requests.
%}
As shown in the input encoding breakdown 
(Figure~\ref{fig:IE_brkdn}), other relatively simple compute operations 
also consume significant number of cycles because they have to wait for 
the long scoreboard to resolve the global memory requests corresponding to the grid lookups. \\

%We also analyzed the \MLP kernel in depth. 
%An \MLP consists of fully connected layers followed by 
%non-linear activation functions such as 
%Sigmoid, TanH, ReLU or Softmax.
Table~\ref{tab:util} also presents the 
the percentage utilization of the GPU compute and memory resources 
for the MLP kernel.
Our results show that for MLP kernels as well, the 
memory utilization is higher than compute utilization. 
This is because, for a constant batch size, the compute cost of the fully 
connected networks scales 
quadratically with the width of the network (Compute Cost: $O(M^2)$)
whereas the memory
traffic scales linearly (Memory Cost: $O(M)$). 
For relatively big MLPs with a large number of 
neurons, the quadratic compute cost quickly becomes 
the bottleneck and further optimizing memory traffic has little to no 
performance benefits because the cost would be overshadowed by the quadratic 
compute requirements. 
However, all our \nG applications have tiny MLPs, 
with only 2-4 hidden layers and 64 hidden neurons per layer (Table~\ref{tab:app_struct}). 
In smaller MLPs, the compute cost and the 
linear memory traffic become asymptotically comparable and hence the 
memory traffic cost starts to matter. 
Moreover, the GPUs and the modern processors in general 
usually have larger computational throughput 
than memory bandwidth, which means that, for small number of neurons, the 
memory cost dominates. \\

\begin{figure}
    \centering
    \includegraphics[width=\linewidth]{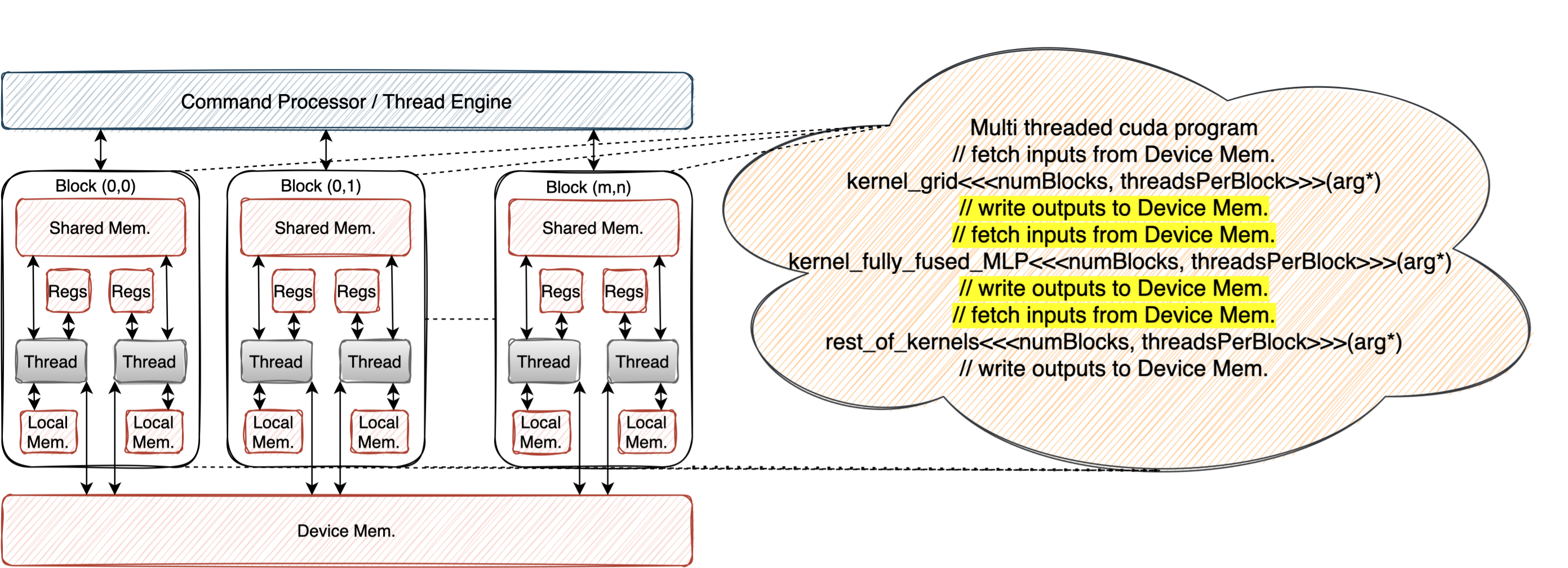}
    \caption{\small Scheduling of a typical neural graphics 
    application on GPU.}
    \label{fig:app_gpu}
    %\vspace{-5mm}
\end{figure}

\iffalse
\begin{figure}
    \centering
    \includegraphics[width=\linewidth]{mlp_on_gpu.drawio.png}
    \caption{Scheduling of \MLP kernel on GPU.}
    \label{fig:mlp_gpu}
\end{figure}
\fi

%\begin{itemize}
%    \item IE analysis 
%    \item F-a utilization table 
%    \item F-b IE bar/pie chart for operations breakdown
%    \item MLP analysis
%    \item F-c MLP parameters for diff apps 
%    \item F-d perf breakdown for one or all the mlps
%    \item F-e utilization table 
%    \item F-f NG application on GPU show R-W-R-W
%\end{itemize}

% \usepackage{booktabs}

\begin{table}[]
\tiny
\centering
\caption{\small The parameters of the four \nG applications for 
different input encoding schemes.}
\label{tab:app_struct}
\begin{tabular}{@{}|c|c|@{}}
\toprule
Application               & Parameters                                                                                                                                                                                                                                                                                                                                   \\ \midrule
NeRF multi res. hashgrid  & \begin{tabular}[c]{@{}c@{}}GridEncoding: Nmin=16 b=1.51572 F=2 T=2\textasciicircum{}19 L=16 \\ Density model: 3--{[}HashGrid{]}--\textgreater{}32--{[}FullyFusedMLP(neurons=64;layers=3){]}--\textgreater{}1 \\ Color model: 3--{[}Composite{]}--\textgreater{}16+16--{[}FullyFusedMLP(neurons=64;layers=4){]}--\textgreater{}3\end{tabular} \\ \midrule
NeRF multi res. densegrid & \begin{tabular}[c]{@{}c@{}}GridEncoding: Nmin=16 b=1.405 F=2 T=2\textasciicircum{}19 L=8 \\ Density model: 3--{[}DenseGrid{]}--\textgreater{}16--{[}FullyFusedMLP(neurons=64;layers=3){]}--\textgreater{}1 \\ Color model: 3--{[}Composite{]}--\textgreater{}16+16--{[}FullyFusedMLP(neurons=64;layers=4){]}--\textgreater{}3\end{tabular}   \\ \midrule
NeRF low res. densegrid   & \begin{tabular}[c]{@{}c@{}}GridEncoding: Nmin=128 b=1 F=8 T=2\textasciicircum{}19 L=2 \\ Density model: 3--{[}TiledGrid{]}--\textgreater{}16--{[}FullyFusedMLP(neurons=64;layers=3){]}--\textgreater{}1 \\ Color model: 3--{[}Composite{]}--\textgreater{}16+16--{[}FullyFusedMLP(neurons=64;layers=4){]}--\textgreater{}3\end{tabular}      \\ \midrule
NSDF multi res. hashgrid   & \begin{tabular}[c]{@{}c@{}}GridEncoding: Nmin=16 b=1.38191 F=2 T=2\textasciicircum{}19 L=16 \\ Model: 3--{[}HashGrid{]}--\textgreater{}32--{[}FullyFusedMLP(neurons=64;layers=4){]}--\textgreater{}1\end{tabular}                                                                                                                            \\ \midrule
NSDF multi res. densegrid  & \begin{tabular}[c]{@{}c@{}}GridEncoding: Nmin=16 b=1.405 F=2 T=2\textasciicircum{}19 L=8 \\ Model: 3--{[}DenseGrid{]}--\textgreater{}16--{[}FullyFusedMLP(neurons=64;layers=4){]}--\textgreater{}1\end{tabular}                                                                                                                              \\ \midrule
NSDF low res. densegrid    & \begin{tabular}[c]{@{}c@{}}GridEncoding: Nmin=128 b=1 F=8 T=2\textasciicircum{}19 L=2 \\ Model: 3--{[}TiledGrid{]}--\textgreater{}16--{[}FullyFusedMLP(neurons=64;layers=4){]}--\textgreater{}1\end{tabular}                                                                                                                                 \\ \midrule
NVR multi res. hashgrid  & \begin{tabular}[c]{@{}c@{}}GridEncoding: Nmin=16 b=1.275 F=2 T=2\textasciicircum{}19 L=16 \\ Model: 3--{[}HashGrid{]}--\textgreater{}32--{[}FullyFusedMLP(neurons=64;layers=4){]}--\textgreater{}4\end{tabular}                                                                                                                              \\ \midrule
NVR multi res. densegrid & \begin{tabular}[c]{@{}c@{}}GridEncoding: Nmin=16 b=1.405 F=2 T=2\textasciicircum{}19 L=8 \\ Model: 3--{[}DenseGrid{]}--\textgreater{}16--{[}FullyFusedMLP(neurons=64;layers=4){]}--\textgreater{}4\end{tabular}                                                                                                                              \\ \midrule
NVR low res. densegrid   & \begin{tabular}[c]{@{}c@{}}GridEncoding: Nmin=128 b=1 F=8 T=2\textasciicircum{}19 L=2 \\ Model: 3--{[}TiledGrid{]}--\textgreater{}16--{[}FullyFusedMLP(neurons=64;layers=4){]}--\textgreater{}4\end{tabular}                                                                                                                                 \\ \midrule
GIA multi res. hashgrid  & \begin{tabular}[c]{@{}c@{}}GridEncoding: Nmin=16 b=1.25992 F=2 T=2\textasciicircum{}24 L=16 \\ Model: 2--{[}HashGrid{]}--\textgreater{}32--{[}FullyFusedMLP(neurons=64;layers=4){]}--\textgreater{}3\end{tabular}                                                                                                                            \\ \midrule
GIA multi res. densegrid & \begin{tabular}[c]{@{}c@{}}GridEncoding: Nmin=16 b=1.405 F=2 T=2\textasciicircum{}24 L=8 \\ Model: 2--{[}DenseGrid{]}--\textgreater{}16--{[}FullyFusedMLP(neurons=64;layers=4){]}--\textgreater{}3\end{tabular}                                                                                                                              \\ \midrule
GIA low res. densegrid   & \begin{tabular}[c]{@{}c@{}}GridEncoding: Nmin=128 b=1 F=8 T=2\textasciicircum{}24 L=2\\  Model: 2--{[}TiledGrid{]}--\textgreater{}16--{[}FullyFusedMLP(neurons=64;layers=4){]}--\textgreater{}3\end{tabular}                                                                                                                                 \\ \bottomrule
\end{tabular}
\end{table}

% \usepackage{booktabs}
% \usepackage{multirow}
\begin{table}[]
\tiny
\centering
\caption{\small The percentage utilization of the
GPU compute and memory resources,
for the input encoding and the MLP kernels
for all four \nG applications and for different input 
encodings.}
\label{tab:util}
\setlength{\tabcolsep}{3pt}
\begin{tabular}{@{}ccccccc@{}}
\toprule
                              &                                        &                                                                                             &                                                                                            &                                                                           &                                                                                                    &                                                                                                  \\
                              &                                        &                                                                                             &                                                                                            &                                                                           &                                                                                                    &                                                                                                  \\
\multirow{-3}{*}{App.-Kernel} & \multirow{-3}{*}{Grid Size/Block Size} & \multirow{-3}{*}{\begin{tabular}[c]{@{}c@{}}Comp. Util. \\ per kernel \\ call\end{tabular}} & \multirow{-3}{*}{\begin{tabular}[c]{@{}c@{}}Mem. Util. \\ per kernel \\ call\end{tabular}} & \multirow{-3}{*}{\begin{tabular}[c]{@{}c@{}}Kernel \\ Calls\end{tabular}} & \multirow{-3}{*}{\begin{tabular}[c]{@{}c@{}}Comp. Util.\\ avg. across \\ application\end{tabular}} & \multirow{-3}{*}{\begin{tabular}[c]{@{}c@{}}Mem. Util.\\ avg. across\\ application\end{tabular}} \\ \midrule
NeRF multi res. hashgrid      & (3853;16;1)/(512;1;1)                  & \cellcolor[HTML]{E06666}61.73                                                               & \cellcolor[HTML]{E06666}72.85                                                              & 59                                                                        & \cellcolor[HTML]{6AA84F}40.63                                                                      & \cellcolor[HTML]{E06666}72.02                                                                    \\
NeRF MLP                      & (3853;16;1)/(512;1;1)                  & \cellcolor[HTML]{6AA84F}34.3                                                                & \cellcolor[HTML]{E06666}65.2                                                               & 118                                                                       & \cellcolor[HTML]{6AA84F}33.36                                                                      & \cellcolor[HTML]{E06666}63.07                                                                    \\
NSDF multi res. hashgrid       & (1823;16;1)/(512;1;1)                  & \cellcolor[HTML]{E06666}73.08                                                               & \cellcolor[HTML]{6AA84F}43.54                                                              & 256                                                                       & \cellcolor[HTML]{93C47D}15.97                                                                      & \cellcolor[HTML]{6AA84F}30.8                                                                     \\
NSDF MLP                       & (1823;16;1)/(512;1;1)                  & \cellcolor[HTML]{6AA84F}38.13                                                               & \cellcolor[HTML]{E06666}71.74                                                              & 256                                                                       & \cellcolor[HTML]{93C47D}9.76                                                                       & \cellcolor[HTML]{93C47D}18.28                                                                    \\
NVR multi res. hashgrid       & (403;16;1)/(512;1;1)                   & \cellcolor[HTML]{E06666}52.5                                                                & \cellcolor[HTML]{E06666}59.03                                                              & 48                                                                        & \cellcolor[HTML]{93C47D}18.67                                                                      & \cellcolor[HTML]{6AA84F}30.36                                                                    \\
NVR MLP                       & (403;16;1)/(512;1;1)                   & \cellcolor[HTML]{6AA84F}36.51                                                               & \cellcolor[HTML]{E06666}67.01                                                              & 48                                                                        & \cellcolor[HTML]{93C47D}11.51                                                                      & \cellcolor[HTML]{93C47D}21.05                                                                    \\
GIA multi res. hashgrid       & (4050;16;1)/(512;1;1)                  & \cellcolor[HTML]{CC0000}82.87                                                               & \cellcolor[HTML]{E06666}62.23                                                              & 1                                                                         & \cellcolor[HTML]{CC0000}82.87                                                                      & \cellcolor[HTML]{E06666}62.23                                                                    \\
GIA MLP                       & (4050;16;1)/(512;1;1)                  & \cellcolor[HTML]{6AA84F}39.1                                                                & \cellcolor[HTML]{E06666}72.22                                                              & 1                                                                         & \cellcolor[HTML]{6AA84F}39.1                                                                       & \cellcolor[HTML]{E06666}72.22                                                                    \\
NeRF multi res. densegrid     & (3966;8;1)/(512;1;1)                   & \cellcolor[HTML]{E06666}71.39                                                               & \cellcolor[HTML]{CC0000}91.81                                                              & 45                                                                        & \cellcolor[HTML]{E06666}57.37                                                                      & \cellcolor[HTML]{E06666}72.31                                                                    \\
NeRF MLP                      & (3966;8;1)/(512;1;1)                   & \cellcolor[HTML]{6AA84F}39.53                                                               & \cellcolor[HTML]{E06666}68.4                                                               & 90                                                                        & \cellcolor[HTML]{6AA84F}34.51                                                                      & \cellcolor[HTML]{E06666}62.31                                                                    \\
NSDF multi res. densegrid      & (1823;8;1)/(512;1;1)                   & \cellcolor[HTML]{CC0000}76.1                                                                & \cellcolor[HTML]{6AA84F}48.25                                                              & 244                                                                       & \cellcolor[HTML]{93C47D}18.38                                                                      & \cellcolor[HTML]{93C47D}21.28                                                                    \\
NSDF MLP                       & (1823;8;1)/(512;1;1)                   & \cellcolor[HTML]{6AA84F}41.66                                                               & \cellcolor[HTML]{E06666}73.49                                                              & 244                                                                       & \cellcolor[HTML]{93C47D}11.06                                                                      & \cellcolor[HTML]{93C47D}19.41                                                                    \\
NVR multi res. densegrid      & (403;8;1)/(512;1;1)                    & \cellcolor[HTML]{E06666}57.38                                                               & \cellcolor[HTML]{E06666}56.8                                                               & 48                                                                        & \cellcolor[HTML]{93C47D}17.41                                                                      & \cellcolor[HTML]{93C47D}22.43                                                                    \\
NVR MLP                       & (403;8;1)/(512;1;1)                    & \cellcolor[HTML]{6AA84F}39.83                                                               & \cellcolor[HTML]{E06666}67.67                                                              & 48                                                                        & \cellcolor[HTML]{93C47D}12.17                                                                      & \cellcolor[HTML]{93C47D}20.59                                                                    \\
GIA multi res. densegrid      & (4050;8;1)/(512;1;1)                   & \cellcolor[HTML]{CC0000}78.53                                                               & \cellcolor[HTML]{E06666}65.83                                                              & 1                                                                         & \cellcolor[HTML]{CC0000}78.53                                                                      & \cellcolor[HTML]{E06666}65.83                                                                    \\
GIA MLP                       & (4050;8;1)/(512;1;1)                   & \cellcolor[HTML]{6AA84F}42.89                                                               & \cellcolor[HTML]{E06666}73.07                                                              & 1                                                                         & \cellcolor[HTML]{6AA84F}42.89                                                                      & \cellcolor[HTML]{E06666}73.07                                                                    \\
NeRF low res. densegrid       & (3980;2;1)/(512;1;1)                   & \cellcolor[HTML]{E06666}53.83                                                               & \cellcolor[HTML]{6AA84F}49.74                                                              & 43                                                                        & \cellcolor[HTML]{6AA84F}31.17                                                                      & \cellcolor[HTML]{E06666}59.57                                                                    \\
NeRF MLP                      & (3980;2;1)/(512;1;1)                   & \cellcolor[HTML]{6AA84F}39.41                                                               & \cellcolor[HTML]{E06666}68.17                                                              & 86                                                                        & \cellcolor[HTML]{6AA84F}35.5                                                                       & \cellcolor[HTML]{E06666}64.1                                                                     \\
NSDF low res. densegrid        & (1823;2;1)/(512;1;1)                   & \cellcolor[HTML]{E06666}55.88                                                               & \cellcolor[HTML]{6AA84F}45.52                                                              & 260                                                                       & \cellcolor[HTML]{93C47D}7.21                                                                       & \cellcolor[HTML]{93C47D}20.07                                                                    \\
NSDF MLP                       & (1823;2;1)/(512;1;1)                   & \cellcolor[HTML]{6AA84F}41.37                                                               & \cellcolor[HTML]{E06666}72.98                                                              & 260                                                                       & \cellcolor[HTML]{93C47D}10.34                                                                      & \cellcolor[HTML]{93C47D}18.14                                                                    \\
NVR low res. densegrid        & (403;2;1)/(512;1;1)                    & \cellcolor[HTML]{93C47D}22.71                                                               & \cellcolor[HTML]{E06666}69.16                                                              & 48                                                                        & \cellcolor[HTML]{93C47D}6.29                                                                       & \cellcolor[HTML]{93C47D}22.71                                                                    \\
NVR MLP                       & (403;2;1)/(512;1;1)                    & \cellcolor[HTML]{6AA84F}39.2                                                                & \cellcolor[HTML]{E06666}66.58                                                              & 48                                                                        & \cellcolor[HTML]{93C47D}12.11                                                                      & \cellcolor[HTML]{93C47D}20.48                                                                    \\
GIA low res. densegrid        & (4050;2;1)/(512;1;1)                   & \cellcolor[HTML]{E06666}66.15                                                               & \cellcolor[HTML]{E06666}59.12                                                              & 1                                                                         & \cellcolor[HTML]{E06666}66.15                                                                      & \cellcolor[HTML]{E06666}59.12                                                                    \\
GIA MLP                       & (4050;2;1)/(512;1;1)                   & \cellcolor[HTML]{6AA84F}42.87                                                               & \cellcolor[HTML]{E06666}73.02                                                              & 1                                                                         & \cellcolor[HTML]{6AA84F}42.87                                                                      & \cellcolor[HTML]{E06666}73.02                                                                    \\ \bottomrule
\end{tabular}
    %\vspace{-0.2in}
\end{table}

\section{Neural Fields Processor: An Architecture for Neural Graphics}
\label{sec:arch}
\iffalse

Our breakdown analysis of the \nG applications (presented in Section~\ref{sec:background})
shows that 
for \ieOne encoding, on average across all four \nG applications \iePercentHashgrid percent of total 
cycles are consumed by the \IE kernel
and \mlpPercentHashgrid percent of total cycles are consumed by the \MLP kernel. 
For \ieTwo encoding, on average across all four \nG applications \iePercentDensegrid percent of total 
cycles are consumed by the \IE kernel
and \mlpPercentDensegrid percent of total cycles are consumed by the \MLP kernel. 
Similarly, for \ieThree \IE, on average across all four \nG applications \iePercentDensegridLowRes percent of total 
cycles are consumed by the \IE
kernel and \mlpPercentDensegridLowRes percent of total cycles are consumed by the \MLP kernel. 
Hence, the \IE and the \MLP are the two most expensive 
kernels across all four representative neural 
graphics applications. This is expected behaviour, as the algorithmic 
analysis of the \nG applications, explained in the section~\ref{sec:background},
suggests that the input encoding and the \MLP 
are the two major building blocks of the \nG applications. 
\fi 

\begin{figure*}
    \centering
    \includegraphics[width=\linewidth]{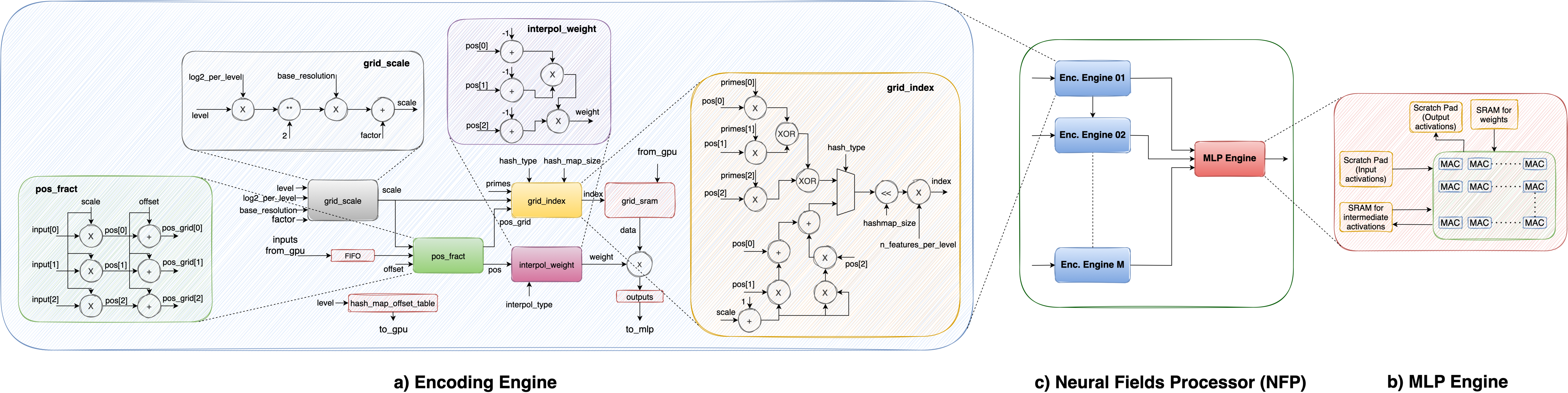}
    \caption{\small Architecture of the Neural Fields Processor 
    (NFP)}
    \label{fig:arch}
\end{figure*}

Since our profiling motivates the need to accelerate 
the \IE and \MLP kernels, we design an architecture 
 - a {\em  \NFP } (Figure~\ref{fig:arch}) - to accelerate  
the \IE and \MLP kernels in hardware using separate hardware engines for the two. 
%\textcolor{red}{
To accelerate \IE kernel, our kernel level analysis (Section~\ref{sec:understandingPerf})
suggests providing hardware support for 
%modulo operations, 
the grid lookups and the dataflow of the input encoding.
%} 
Moreover, in neural graphics applications, the outputs of the \IE kernel
are always consumed by the \MLP kernel (Figure ~\ref{fig:all_app_struct}) 
%\textcolor{red}{
In 
%Nvidia's 
Müller et al's
GPU implementation
of these applications~\cite{ngpgithub, muller2022instant},
%} 
the \IE 
kernel writes outputs to device memory and 
the \MLP kernel fetches that data again for 
further processing (Figure~\ref{fig:app_gpu}). 
This leads to unnecessary memory traffic and 
waste of energy for DRAM accesses that can 
potentially be avoided. 
This opportunity can be exploited in the \NFP 
hardware by fusing the \IE and \MLP 
engines in such a way that the \IE engine 
directly writes the outputs to the input memory 
of the \MLP engine. \\

%In order to accelerate the two most 
%expensive
%primitives of the \nG applications (\IE and 
%\MLP), we present a {\em  \NFP } in 
%Figure~\ref{fig:arch}.
%The \NFP architecture 
%is inspired by the insights learned from the
%breakdown analysis of the \nG applications. 
%Each \NFP has an \IE IP and an \MLP 
%engine fused together in hardware. 
Figure~\ref{fig:arch}-a presents 
the architecture of the \IE hardware engine. 
Each \IE hardware engine has a dedicated 
on-chip SRAM (\emph{grid\_sram}) to cache the 
lookup 
table for 
one resolution level.
The size of the \emph{grid\_sram} 
(1MB per \IE engine)
is chosen such that 
the entire lookup table for one resolution
level fits on the on-chip SRAM, and the 
off-chip memory access penalty
could be avoided for grid lookups. 
The lookup table for one resolution level is 
cached once on the dedicated 
\emph{grid\_sram} of one \IE engine and then lookups 
are performed for all the inputs 
for the entire frame. 
Our hardware architecture has 16
\IE engines to match the maximum number of 
resolution 
levels of the \nG applications. 
As the \ieOne \IE has 16 resolution levels,
each of the 16 \IE engines caches the lookup table 
corresponding to one resolution level and then
performs the \IE for all the resolution 
levels in parallel.
As \ieTwo \IE has 8 resolution levels, 
two inputs can be processed in parallel. 
Similarly, as \ieThree \IE has 2 resolution 
levels, the \IE IP can process 8 inputs 
in parallel. 

The normalized input coordinates 
are pre-fetched into 
the input FIFO (Figure~\ref{fig:arch}-a).
The \emph{grid\_scale} module calculates the 
scale of the grid from the base-resolution of 
the grid and the resolution 
level being processed. 
The \emph{pos\_fract} module is responsible 
for converting the normalized input 
coordinates to the absolute 
coordinates. It multiplies the 
normalized coordinates with grid scale 
to compute the absolute coordinates, which are
then
passed to the
\emph{grid\_index} module for the index 
calculation. 
The \emph{grid\_index} module is responsible for 
calculating the final indices for 
the lookup operations.
The 
\emph{grid\_index} module can be 
configured to either hash in the indices
for the \ieOne encoding or compute the indices 
without the hashing function for the \ieTwo
and \ieThree encoding types. 
The algorithm to compute the indices for the 
lookup operation
takes the 
modulo of the indices with the 
hash-map size
as an intermediate operation 
before the final indices could be calculated.
We observe that the hash-map size 
is always power of two for all our \nG
applications. 
We exploit this optimization opportunity in 
hardware and approximate the modulo operation 
with shift operation in \IE engine. 
The final outputs of the \emph{grid\_index} 
modules are actual indices that are directly 
used to perform the feature lookups from the 
\emph{grid\_sram}. The \emph{interpol\_weights}
module 
computes the interpolation weights that are then
multiplied with the features to compute the 
final features that are fed to the \MLP engine. 
%The data-flow and the corresponding compute 
%operations of each module of the \IE engine
%are shown in Figure~\ref{fig:arch}-a.
As each \IE engine calculates features for one 
resolution level, the outputs of the 
\IE engines are concatenated together to get the 
final input vector for the MLP engine. 

Since MLPs in \nG applications are small (2-4 hidden layers,  64 neurons in each hidden layer), 
%For architecture of the MLP engine, consider
%the structure of the \MLP for 
%the \nG applications (Table~\ref{tab:app_struct}). 
%We observe that
%the \MLP in all the \nG applications have 
%2-4 hidden layers and the number of hidden 
%neurons is always 64 in each hidden layer of the 
%\MLP.
%This is because part of the learning 
%task is offloaded to the learnable 
%parameters of \IE and relatively 
%smaller \MLP with only 2-4 hidden layers and 
%small number of hidden neurons (64 hidden 
%neurons per layer) are enough to synthesize good 
%quality output images. 
%Based on this observation, 
our MLP engine has a 
$64\times64$ grid of MAC units that computes 
one layer of the \MLP at a time.
As the size of the hidden layer is relatively 
small, a dedicated small on-chip SRAM is
used to store the 
intermediate features of the hidden layers. 
Keeping the intermediate features on-chip 
removes the off-chip memory accesses 
for storing/fetching the intermediate features 
and improves the performance by 1OOM 
~\cite{muller2021real}. \\

\begin{figure*}
    \centering
    \includegraphics[width=\linewidth]{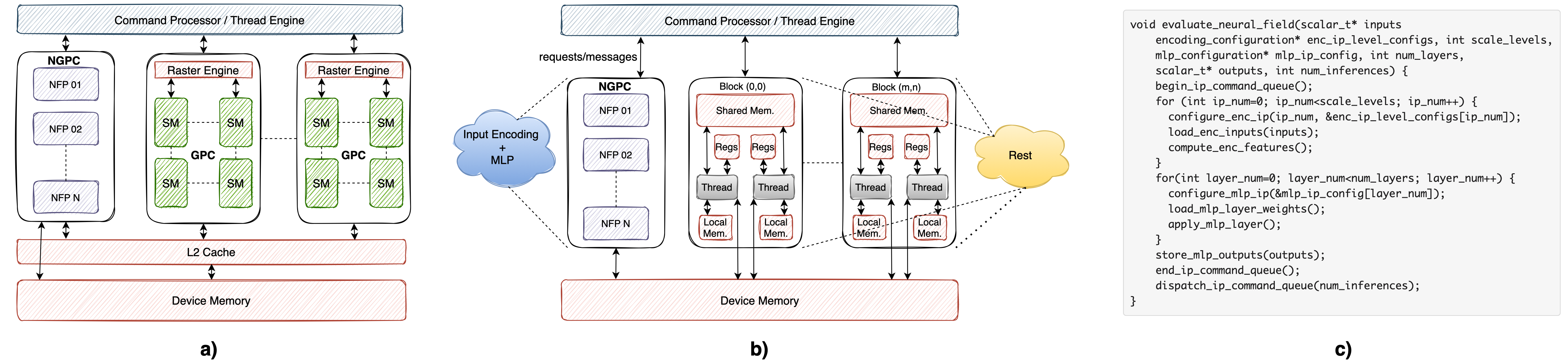}
    \caption{\small a) Interaction of the \npc{} with GPU.
    b) Programming model for \npc{}. c) Pseudocode for programming \npc{}.}
    \label{fig:gpu_interact}
        \vspace{-0.2in}
\end{figure*}

Figure~\ref{fig:gpu_interact}-a shows the 
interaction of the \NFP with the GPU. 
%A conventional Nvidia GPU has Graphics 
%processing clusters (GPCs) that have multiple 
%streaming processors (SMs) connected to a 
%shared L2 cache. 
%We $\nfp$s  scalable \NPC (\npc) along with 
%the existing GPC units. The proposed 
A set of N NFPs are organized as a \NPC (\npc{}) 
connected to the shared L2 cache.  
%An \npc can 
%be scaled to have N number of \nfp units. 
Figure~\ref{fig:gpu_interact}-b 
and~\ref{fig:gpu_interact}-c show the 
programming model of the \npc{}. 
%The command processor 
%(Giga Thread Engine in Nvidia's terminology)
%schedules the CUDA kernel on the streaming multiprocessors by 
%clustering the streaming multiprocessors
%in blocks.
%We propose a 
The programming model 
for the \npc{} involves  the GPU 
command buffer~\cite{commandBuffer}
configuring the \npc{} and scheduling the 
\IE and the \MLP kernels on the \npc{}.
The rest of the kernels are scheduled 
just as conventional CUDA kernels 
on the streaming multiprocessors.
The outputs of the \npc{} are written back 
to the GPU memory, and are read by the 
streaming multiprocessors, which then compute 
the rest of the kernels.  
The inputs are divided into batches.
While the GPU is processing the rest of the 
\nG application kernels for the Nth batch of 
inputs, the N+1st batch is scheduled on the 
\npc{} to compute the \IE and MLP
kernels in parallel, as shown in 
Figure~\ref{fig:gpu_interact}-b. 

%\begin{itemize}
%    \item arch F-a arch [table:IE ip steady state]
%    \item interaction F-b interaction
%    \item programming 
%    \item F-c programming multiple IPs multiple copies
    
%\end{itemize}

\section{Evaluation and Results}
\label{sec:results}
\begin{figure}
    \centering
    \includegraphics[width=\linewidth]{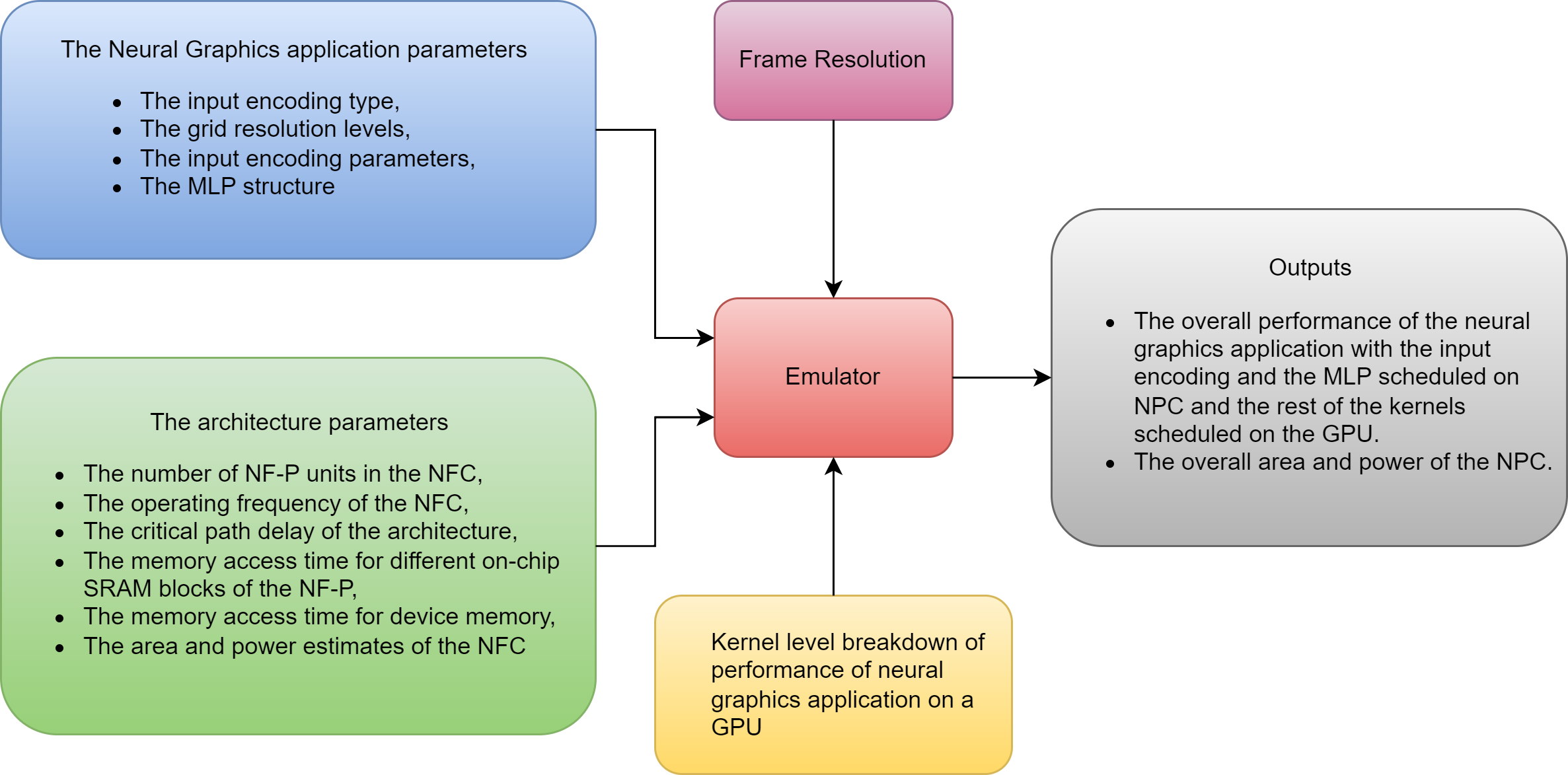}
    \caption{Block diagram of the emulator.}
    \label{fig:emulator}
\end{figure}

We provide an emulator to evaluate the 
performance 
of the \nG applications on our architecture. 
The block diagram of the emulator is 
in Figure~\ref{fig:emulator}. 
The inputs to the emulator are 1) The \nG 
application parameters such as the \IE type, 
the grid resolution levels, the \IE parameters, 
the \MLP structure 2) the architecture parameters
such as the number of \nfp units in the 
\npc{}, operating frequency of the \npc{}, 
the critical path delay of the architecture, 
the memory access time for different on-chip
SRAM blocks of the \nfp, the memory access time
for device memory, the area and power estimates 
of the \npc{} 
3) the kernel level breakdown of the 
performance of 
the \nG application on the GPU
and 4) the frame resolution. 
The outputs of the emulator are 
1) the overall performance of the 
\nG application with the \IE and the \MLP
scheduled on \npc{} and the rest of the 
kernels scheduled on the GPU. 
2) the overall area and power of the 
\npc{}. 

\begin{figure*}[ht!]
   \subfloat[\label{fig:app_perf_mrhg}]{%
      \includegraphics[width=0.32\textwidth]{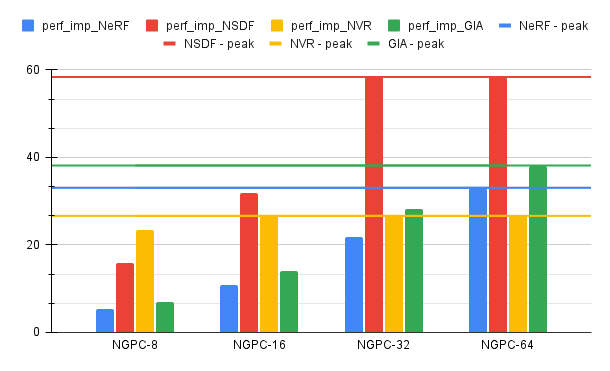}}
\hspace{\fill}
   \subfloat[\label{fig:app_perf_mrdg} ]{%
      \includegraphics[width=0.32\textwidth]{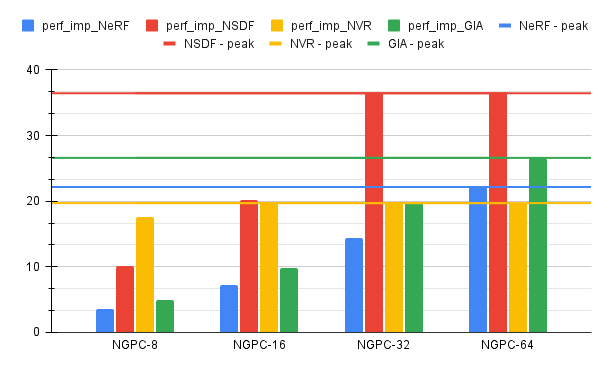}}
\hspace{\fill}
   \subfloat[\label{fig:app_perf_srdg}]{%
      \includegraphics[width=0.32\textwidth]{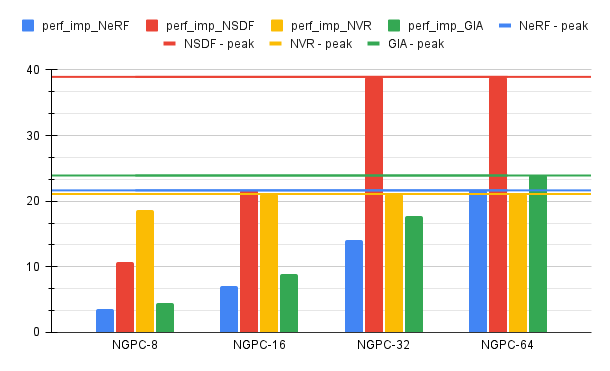}}\\
\caption{The performance of the neural graphics 
applications on our proposed architecture for the scaling 
factor of 8, 16, 32 and 64. 
NGPC-8 has 8 NFP units, NGPC-16 has 16 NFP units and so on.
(a) {\em \iEOne encoding}, 
(b) {\em \iETwo encoding}, 
(c) {\em \iEThree encoding}.}
\end{figure*}

We evaluated the performance of the 
\nG applications on our proposed architecture
for the scaling factor of 8, 16, 32 and 64, where
NGPC-8 has 8 \nfp unit, NGPC-16 has 16 \nfp units
and so on.
Figure~\ref{fig:app_perf_mrhg} presents 
the overall performance of the \nG applications 
with \ieOne encoding on the proposed architecture.
%and \MLP scheduled 
%on \npc and the rest of the kernels 
%scheduled on the GPU. 
Our results show that when we have only 
8 \nfp units in the NGPC,
on average across all four 
\nG applications, we get overall 
$\npcOAvgHG\times$
performance benefits compared to the GPU baseline. 
%case where the entire \nG application is 
%scheduled on the streaming multiprocessors of the 
%GPU.
When we scale the number of \nfp units 
in an \npc{} to 16, 32 and 64,
%compared to the baseline case, 
the performance 
benefits 
averaged across the four representative \nG 
applications increase to 
$\npcTAvgHG\times$, $\npcFAvgHG\times$ and 
$\npcEAvgHG\times$ 
respectively.

Figure~\ref{fig:app_perf_mrdg} presents 
the overall performance of the \nG applications 
with \ieTwo encoding.
%and \MLP scheduled 
%on \npc and the rest of the kernels 
%scheduled on the GPU. 
Our results show that when we have only 
8 \nfp units in \npc{}, 
on average across all four 
\nG applications, we get overall 
$\npcOAvgDG\times$
performance benefits compared to the GPU baseline.
%where the entire \nG application is 
%schedule on the streaming multiprocessors of 
%the GPU.
When we scale the number of \nfp units 
in an \npc{} to 16, 32 and 64,
%compared to the baseline case, 
the performance 
benefits 
averaged across the four representative \nG 
applications increase to 
$\npcTAvgDG\times$, $\npcFAvgDG\times$ and 
$\npcEAvgDG\times$ 
respectively.

Figure~\ref{fig:app_perf_srdg} presents 
the overall performance of the \nG applications 
with \ieThree encoding.
%and \MLP scheduled 
%on \npc and the rest of the kernels 
%scheduled on the GPU. 
Our results show that when we have only 
8 \nfp units in \npc{}, 
on average across all four 
\nG applications, we get overall 
$\npcOAvgLR\times$
performance benefits compared to the GPU baseline.
%where the entire \nG application is 
%schedule on the streaming multiprocessors of 
%the GPU.
When we scale the number of \nfp units 
in an \npc{} to 16, 32 and 64, 
%compared to the baseline case, 
the performance 
benefits 
averaged across the four representative \nG 
applications increase to 
$\npcTAvgLR\times$, $\npcFAvgLR\times$ and 
$\npcEAvgLR\times$ 
respectively.

Our results also show that
\appN  
performance plateaus for 
\npc{}-64. I.e., increasing the number of \nfp 
beyond 64 does not improve the overall 
performance of the application. This 
is because the 
time consumed by the non- \IE and \MLP kernels
%GPU for performing the 
%rest of the kernels of the \appN 
becomes the performance bottleneck. 
Similarly, for \appS, \appV and \appI the
performance plateaus for \npc{}-32, \npc{}-16 and 
\npc{}-64 respectively.

\begin{figure*}[ht!]
   \subfloat[\label{fig:indv_perf_mrdg}]{%
      \includegraphics[width=0.32\textwidth]{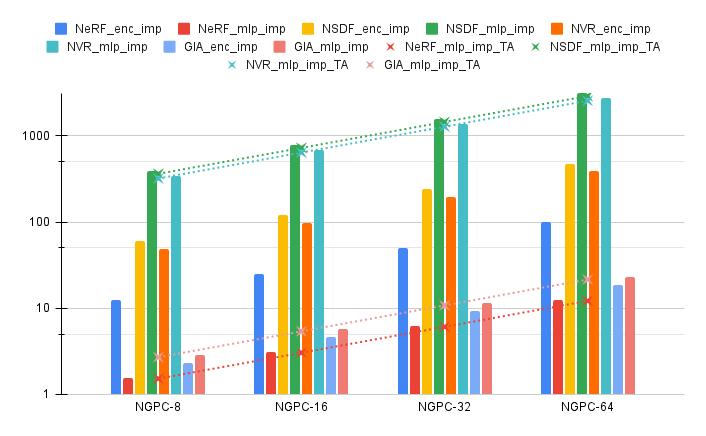}}
\hspace{\fill}
   \subfloat[\label{fig:indv_perf_mrdg} ]{%
      \includegraphics[width=0.32\textwidth]{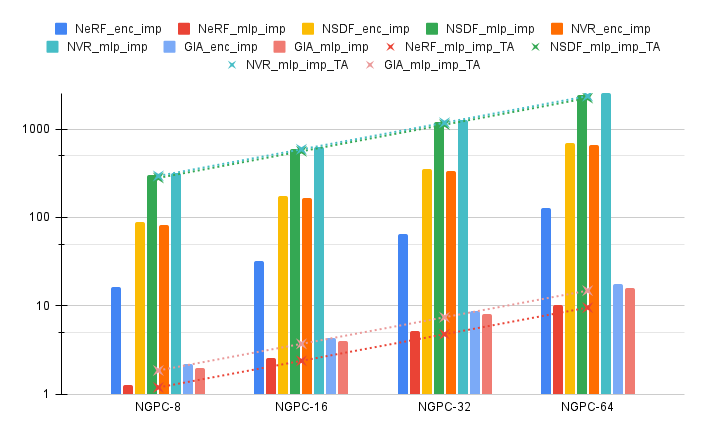}}
\hspace{\fill}
\subfloat[\label{fig:indv_perf_sr} ]{%
      \includegraphics[width=0.32\textwidth]{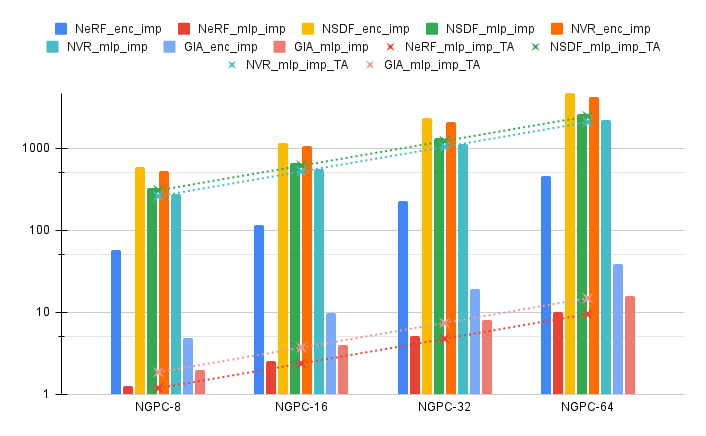}}
%\hspace{\fill}
%   \subfloat[\label{fig:indv_perf_avg}]{%
%      \includegraphics[width=0.23\textwidth]{indv_perf.png}}\\
\caption{\small The performance improvement of the input 
encoding kernels and 
the multi-layer perceptron kernels 
individually, on our architecture, for the 
scaling factors of 8, 16, 32 and 64. For all four \nG applications and different input encoding types.
(a) {\em \iEOne encoding},
(b) {\em \iETwo encoding},
(c) {\em \iEThree encoding},
%(d) average across all four \nG applications 
where dotted lines labeled as mlp\_imp\_TA are the performance improvements 
of MLP engine modeled with timeloop and accelergy.}
\label{fig:indv_perf}
\end{figure*}

In order to better understand where the application level
benefits are coming from, we also compare the performance of the
\IE and the \MLP kernels individually,
with their GPU based implementations. 
Figure~\ref{fig:indv_perf} presents the performance improvement of the
\IE kernel and the \MLP kernels individually,
on our architecture, for scaling factors of 8, 16, 32 and 64. 
Our results show that 
for \ieOne, on average across four \nG applications, the 
\NPC-64 has performance improvement of 
$\hgIeIndv \times$ and $\hgMlpIndv \times$
for the \IE kernel and the \MLP kernel, respectively. 
For \ieTwo,
%, on average across four \nG applications, the 
\NPC-64 has performance improvement of 
$\dgIeIndv \times$ and $\dgMlpIndv \times$
for the \IE kernel and the \MLP kernel, respectively. 
For \ieThree, 
%on average across four \nG applications, 
the 
\NPC-64 has performance improvement of 
$\srIeIndv \times$ and $\srMlpIndv \times$
for the \IE kernel and the \MLP kernel, respectively.

\begin{figure*}[ht!]
   \subfloat[\label{fig:fps_mrdg}]{%
      \includegraphics[width=0.32\textwidth]{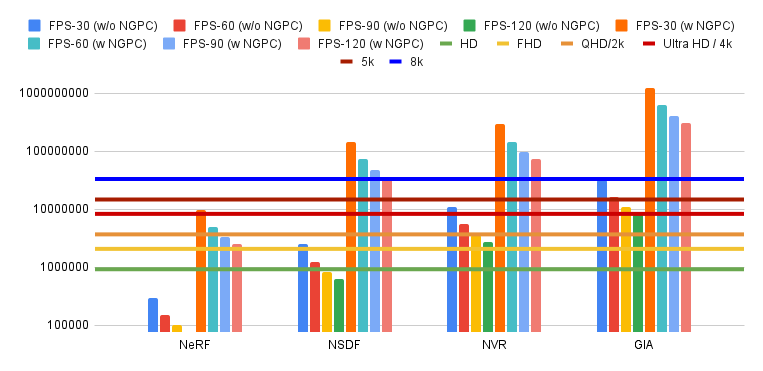}}
\hspace{\fill}
   \subfloat[\label{fig:fps_mrdg} ]{%
      \includegraphics[width=0.32\textwidth]{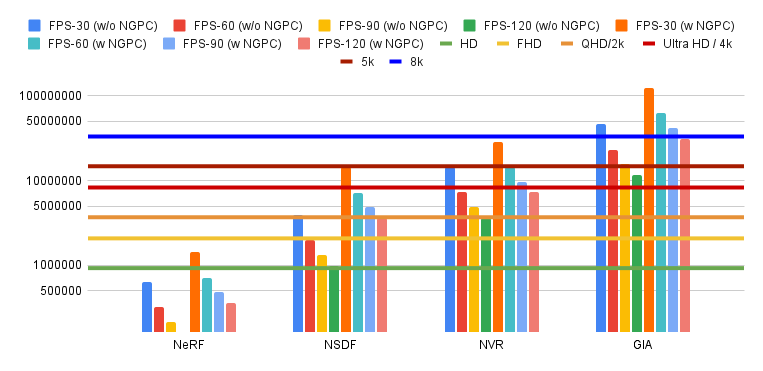}}
\hspace{\fill}
   \subfloat[\label{fig:fps_sr}]{%
      \includegraphics[width=0.32\textwidth]{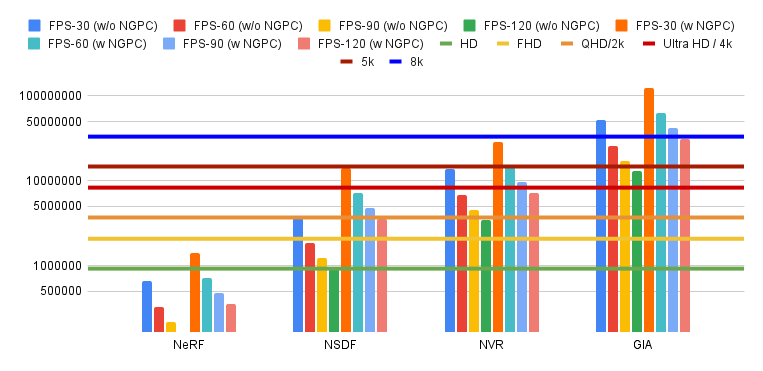}}\\
\caption{\small The number of pixels that can be
rendered for a given FPS target with and without neural 
graphics processing cluster. 
(a) {\em \iEOne encoding}, 
(b) {\em \iETwo encoding}, 
(c) {\em \iEThree encoding}. }
\label{fig:fps}
\end{figure*}

In Figure~\ref{fig:fps}, 
we present the number of pixels that 
can be rendered for a given FPS target with and without \NPC.
Horizontal lines in the 
figure mark the number of pixels in HD ($1280\times720$), FHD ($1920\times1080$), 
QHD/2k ($2560\times1440$), Ultra HD/4k ($3820\times2160$), 5k ($5120\times2880$) and 
8k ($7680\times4320$) frame resolutions.
The vertical bars show the number of pixels rendered 
within the time budget of 1000/30 = 33.33ms, 1000/60 = 16.67ms, 1000/90 = 11.11ms and 1000/120 = 8.33ms
corresponding to the FPS targets of 30, 60, 90 
and 120 FPS
respectively.
Our results show that with \ieOne encoding, 
\npc{} enables the rendering of 
4k Ultra HD resolution frames at 30 FPS for NeRF and
8k Ultra HD resolution frames at 120 FPS for all our 
other neural graphics applications.
%30 FPS HD frames for NeRF,
%20 FPS QHD (2k resolution) frames
%and 30 FPS 5k resolution frames for NSDF,
%60 FPS 5k resolution,
%90 FPS Ultra HD (4k resolution) 
%and 
%120 FPS QHD (2k resolution)
%frames for NVR,
%and 120 FPS 5k resolution
%frames for GIA.

\begin{figure}
    \centering
    \includegraphics[width=0.75\linewidth]{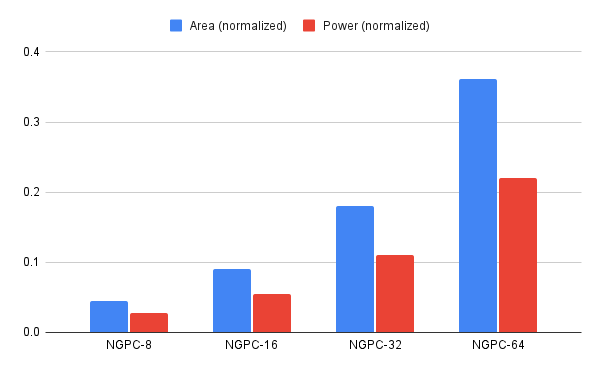}
    \caption{\small The area and Power of \NPC, normalized with 
    respect to the area and power of Nvidia RTX 3090 die area and 
    power.}
    \label{fig:norm_ap}
    %\vspace{-5mm}
\end{figure}

For estimating the area and power 
overheads,
%of the \NPC, 
we wrote RTL for \NFP and synthesized it 
using Synopsys design compiler along with 
the Nangate 45nm open cell library.
We used CACTI to get the area and 
power estimates for the SRAM blocks.
%of the 
%\nfp.
%As our proposed architecture 
%sits on a GPU die~\ref{fig:arch},
%we also estimated the area and power overhead of 
%putting the \npc on GPU.
%compare the area and power of \npc 
%with GPU area and power numbers. 
Figure~\ref{fig:norm_ap} shows the 
area and power of the different configurations of 
\npc{} normalized with 
respect to Nvidia's RTX3090 area and power.
In order to get iso-technode comparison, we 
scaled the area and power of \npc{} to 7nm 
using often-used scaling formulas~\cite{stillmaker2017scaling}.
Our estimates show that \npc{}-8 
that has only one \nfp unit 
increases the die area of GPU by only 
$\sim\percentOarea\%$ with the power overhead of 
$\sim\percentOpow\%$. 
Similarly, \npc{}-16, \npc{}-32 and \npc{}-64
increase the GPU die area by 
$\sim\percentTarea\%$, 
$\sim\percentFarea\%$ and 
$\sim\percentEarea\%$
respectively and GPU power by 
$\sim\percentTpow\%$, 
$\sim\percentFpow\%$ and 
$\sim\percentEpow\%$
respectively.

Table~\ref{tab:bw_npc} presents the input/output bandwidth
and the data access time for our \npc{} architecture.
Our estimates suggest that, 
for 60FPS, the bandwidth requirement for \npc{} architecture
is 231 GB/s for NeRF and 69 GB/s for all other \nG applications. 
The memory bandwidth of Nvidia RTX 3090 is 936.2 GB/s ~\cite{gpubw}. 
Hence, for 60FPS, 
the IO bandwidth of the accelerator is $\sim24\%$
of the GPU memory bandwidth for NeRF and only $\sim7\%$
of the GPU memory bandwidth for NSDF, NVR and GIA. 
Moreover, the massively parallel nature of the workload and 
the high memory bandwidth of the GPU compared to the IO 
bandwidth of the accelerator keeps the encoding engines busy 
with high utilization. 
This translates to data access time of 4.12ms for NeRF and 
1.23ms for all other \nG applications.

In order to get confidence in our 
evaluation methodology and the reported speedup numbers, we performed a sanity check against 
Amdahl's law and presented our analysis 
in figures~\ref{fig:app_perf_mrhg},~\ref{fig:app_perf_mrdg} 
and~\ref{fig:app_perf_srdg}. Horizontal lines in
the figures~\ref{fig:app_perf_mrhg},~\ref{fig:app_perf_mrdg} 
and~\ref{fig:app_perf_srdg} show the peak speedup bounded by Amdahl's law 
and the vertical bars show the reported speedup from our emulator. 
Our analysis shows that the reported speedup 
is always under the Amdahl-driven analytical bounding of speedup.
We also modeled the performance of our MLP engine using
popular open-source DNN-architecture-modeling frameworks
Timeloop~\cite{parashar2019timeloop} and 
Accelergy~\cite{wu2019accelergy}.
In figure~\ref{fig:indv_perf} we also present the performance
benefits of the MLP engine modeled with the 
timeloop and accelergy.
Our analysis shows that the performance benefits reported 
by our emulator are within $\sim7\%$ of the the performance 
benefits modeled with the timeloop and accelergy.

%After the design space 
%exploration of our architecture 
%for the scaling factors of 1, 2, 4 and 
%8. We propose to build 
%\npc with 8 \nfp units to get maximum benefits 
%for all four \nG applications.}

%\begin{itemize}
%    \item app x IE results
%    \item deeper analysis results breakdown
%    \item area/power analysis results 
%    \item scalability analysis 
%\end{itemize}

%\begin{figure}
%   \centering
%    \includegraphics[width=\linewidth]{indv_perf.png}
%    \caption{Performance of input encoding engine and the MLP engine normalized with respect to GPU.}
%    \label{fig:indv_perf}
%\end{figure}

% \usepackage{booktabs}
\begin{table}[]
\centering
\tiny
\caption{\small The input/output bandwidth and the data
access time for our NGPC architecture.}
\label{tab:bw_npc}
\begin{tabular}{@{}ccccc@{}}
\toprule
App. & Input BW (GB/s) & Output BW (GB/s) & Totoal BW (GB/s) & Access time (ms) \\ \midrule
NeRF & 69.523          & 46.349           & 231.743          & 4.126              \\
NSDF & 34.761          & 34.761           & 69.523           & 1.238              \\
GIA  & 34.761          & 34.761           & 69.523           & 1.238              \\
NVR  & 34.761          & 34.761           & 69.523           & 1.238              \\ \bottomrule
\end{tabular}
\vspace{-5mm}
\end{table}

\section{Summary and Conclusion}
\label{sec:summaryAndConclusion}
Neural graphics promises a fast, deterministic time replacement for traditional rendering algorithms. In this paper, we address the question:
{\em does neural graphics need hardware support?}
%We studied four representative \nG applications and identified that 
%there is the gap of
%$\sim 2OOM-4OOM$ between
%the desired performance and power and the state of the art.
%\textcolor{red}{
We studied four representative \nG applications (NeRF, NSDF, NVR, and GIA) and 
%identified the performance and power gap between the 
%desired targets and the state of the art.
%Our profiling 
showed that, if we want to render 4k resolution 
frames at 60FPS, there is the gap of 
$\sim\gapfpsVHGF \times$ to $\gapfpsNHGF \times$
between
the desired performance and the state of the art.
For AR and VR applications, 
there is a larger gap of
$\sim$ 2-4OOM 
between
the desired performance and target power.
%}
Through in depth analysis,
%of the 
%\nG applications 
%and 
we identified that the \IE and the \MLP kernels are the
performance bottlenecks
%.
%Our analysis showed that, 
%on averaged across all four representative \nG applications,
consuming
$\ieOneIEMLP \%$, $\ieTwoIEMLP \%$ and $\ieThreeIEMLP \%$ of application time 
%is consumed by the \IE and the \MLP kernels, 
for {\em \ieOne encoding}, {\em \ieTwo encoding} and {\em \ieThree encoding} respectively. 
Based on the compute and memory access characteristics of the 
\IE and the \MLP kernels, 
we proposed \NPC -- a scalable hardware architecture that directly accelerates the \IE and \MLP 
kernels through dedicated engines and supports a wide range of \nG applications.
%We also accelerate the rest of the kernels by fusing them 
%together in Vulkan~\cite{vulkan}, which leads to
%$\sim\fuserest\times$ kernel-level performance improvement compared to 
%Nvidia's "un-fused" implementation~\cite{muller2022instant} of the pre-processing and the post-processing kernels.
%\textcolor{red}{
To achieve good overall application level performance improvements, we also accelerate the rest of the kernels by fusion into a single kernel, leading to a $\sim\fuserest\times$ speedup compared to previous optimized implementations~\cite{muller2022instant} which is sufficient to remove this performance bottleneck.
%}
Our results show that,
\npc{} gives up to $\maxperf\times$ 
end-to-end application-level performance improvement.
For {\em \ieOne encoding} on average across the four \nG applications, 
the performance benefits of our architecture are
$\npcOAvgHG\times$, $\npcTAvgHG\times$, $\npcFAvgHG\times$ and 
$\npcEAvgHG\times$ 
for the scaling factor of 8, 16, 32 and 64, respectively.
For {\em \ieTwo encoding} on average across the four \nG applications, 
the performance benefits of our architecture are
$\npcOAvgDG\times$, $\npcTAvgDG\times$, $\npcFAvgDG\times$ and 
$\npcEAvgDG\times$ 
for the scaling factor of 8, 16, 32 and 64, respectively.
Similarly, for {\em \ieThree encoding} on average across the four \nG applications, 
the performance benefits of our architecture are
$\npcOAvgLR\times$, $\npcTAvgLR\times$, $\npcFAvgLR\times$ and 
$\npcEAvgLR\times$ 
for the scaling factor of 8, 16, 32 and 64, respectively.
Our results show that with \ieOne encoding, 
\npc{} enables the rendering of 
4k Ultra HD resolution frames at 30 FPS for NeRF and
8k Ultra HD resolution frames at 120 FPS for all our 
other neural graphics applications. 
%30 FPS HD frames for NeRF,
%20 FPS QHD (2k resolution) frames
%and 30 FPS 5k resolution frames for NSDF,
%60 FPS 5k resolution,
%90 FPS Ultra HD (4k resolution) 
%and 
%120 FPS QHD (2k resolution)
%frames for NVR,
%and 120 FPS 5k resolution
%frames for GIA.

\section{Acknowledgements}
We thank Selvakumar Panneer (Intel) for multiple discussion 
about
NeRF and neural graphics that helped conceive this project.
We thank
Anton Kaplanyan, Rama Harihara,
Nilesh Jain, Ravi Iyer and Maxim Kazakov (from Intel)
for multiple discussions
about neural graphics. 
We also thank Vikram Sharma Mailthody,
the anonymous reviewers as
well as the members of the Passat group for their feedback.

%%%%%%% -- PAPER CONTENT ENDS -- %%%%%%%%

%%%%%%%%% -- BIB STYLE AND FILE -- %%%%%%%%
\bibliographystyle{IEEEtranS}
\bibliography{refs}
%%%%%%%%%%%%%%%%%%%%%%%%%%%%%%%%%%%%

\end{document}